\documentclass[reprint,superscriptaddress,aps,prl]{revtex4-2}
\usepackage{makeidx}
\usepackage[utf8]{inputenc}
\usepackage{graphicx}
\usepackage{braket}
\usepackage{amsmath}
\usepackage{xcolor}
\usepackage{amssymb}
\usepackage[normalem]{ulem}

\usepackage[breaklinks,colorlinks,urlcolor=blue,linkcolor=blue,anchorcolor=blue,citecolor=blue]{hyperref}
\definecolor{darkgreen}{RGB}{0,160,0}

\def\ga{\Gamma_\text{1D}}  
\newcommand\varpm{\mathbin{\vcenter{\hbox{
  \oalign{\hfil$\scriptstyle+$\hfil\cr
          \noalign{\kern-.3ex}
          $\scriptscriptstyle({-})$\cr}}}}}
\begin{document}
\begin{abstract}
    Computing the exact dynamics of many-body quantum systems becomes intractable as system size grows. Here, we present a symmetry-based method that provides an exponential reduction in the complexity of a broad class of such problems -- qubits coupled to one-dimensional electromagnetic baths. We identify conditions under which partial permutational symmetry emerges and exploit it to group qubits into collective multi-level degrees of freedom, which we term ``superspins.'' These superspins obey a generalized angular momentum algebra, reducing the relevant Hilbert space dimension from exponential to polynomial. Using this framework, we efficiently compute many-body superradiant dynamics in large arrays of qubits coupled to waveguides and ring resonators, showing that -- unlike in conventional Dicke superradiance -- the total spin length is not conserved. At long times, dark states become populated. We identify configurations where these states exhibit metrologically useful entanglement. Our approach enables exact treatment of complex dissipative dynamics beyond the fully symmetric limit and provides a rigorous benchmark for approximate numerical methods.
\end{abstract}

\title{Exact many-body quantum dynamics in one-dimensional baths via collective spins}
\author{Joseph T. Lee}
\thanks{These authors contributed equally to this work.}
\affiliation{Department of Physics, Columbia University, New York, New York 10027, USA}
\author{Silvia Cardenas-Lopez}
\thanks{These authors contributed equally to this work.}
\affiliation{Department of Physics, Columbia University, New York, New York 10027, USA}
\author{Stuart J. Masson}
\affiliation{Department of Physics, University of South Florida, Tampa, Florida 33620, USA}
\author{Rahul Trivedi}
\affiliation{Max Planck Institute of Quantum Optics, Hans-Kopfermann-Str. 1, Garching 85748, Germany}
\author{Ana Asenjo-Garcia}
\email{ana.asenjo@columbia.edu}
\affiliation{Department of Physics, Columbia University, New York, New York 10027, USA}
\maketitle

The exponential growth of the Hilbert space in many-body problems is the main difficulty for their efficient simulation. An important exception arises in systems with a high degree of symmetry, which impose constraints that significantly reduce the relevant Hilbert space size. In open quantum systems, this is exemplified by multiple qubits coupled identically to a shared bath. The full permutational symmetry among the qubits is preserved in the master equation obtained by tracing out the bath degrees of freedom, drastically reducing the complexity of the problem. Specifically, an ensemble of $N$ spin-$\frac{1}{2}$ particles interacting collectively can be described by a single spin-$\frac{N}{2}$ representation. This mapping reduces the relevant Hilbert space dimension from $2^N$ to $N+1$, corresponding precisely to the fully symmetric subspace spanned by the Dicke states~\cite{Dicke54}. This theoretical idealization is realized experimentally in systems of qubits coupled identically to a single mode Fabry-P\'{e}rot cavity~\cite{Ritsch13,Mivehvar21}, and has been harnessed for different applications, including the generation of entangled states of matter~\cite{Hosten16,Cox16,Colombo22,Cooper24}, the study of quantum phase transitions~\cite{Baumann10,Periwal21,Young24}, and the realization of novel photon sources~\cite{Bohnet12,Simon20,Meiser09}.

The use of symmetry to reduce the complexity of open quantum dynamics extends beyond fully permutational scenarios. For example, symmetry-breaking processes such as local incoherent pumping, dephasing, and decay to additional baths have been efficiently treated by introducing collective spin representations with spin lengths smaller than $N/2$~\cite{Sarkar87,Chase08,Molmer18,Alvarez20}. Despite these processes breaking the symmetry locally (at the single-emitter level), permutational symmetry is preserved on average, specifically at the level of the density matrix, provided these processes occur identically for each emitter. Such dynamics are accurately described using a SU(4) operator formalism~\cite{Hartmann16, Holland13}, yielding a polynomial ($\sim N^3$) scaling of the density matrix dimension. Extensions to multilevel spin systems further increase the exponent of this polynomial scaling~\cite{Bolanos15}. More broadly, Lie-algebraic methods have been employed to systematically identify and characterize quantum dynamics that admit efficient classical simulation~\cite{Anschuetz2023, Goh2023}. 

In this work, we demonstrate that the dissipative dynamics of an ordered array of qubits coupled to a one-dimensional (1D), structureless, and lossless electromagnetic bath (e.g. a ring cavity or a single-mode waveguide) can be described efficiently, with computational complexity scaling only polynomially with system size. Our approach leverages the partial permutational symmetry that emerges at specific lattice constants, allowing subsets of qubits to be mapped into collective spin degrees of freedom, which we term ``superspins.'' The emergence of these superspins can be understood either by directly inspecting the structure of the matrix that describes the dissipative coupling between qubits or, equivalently, through the Lie algebra generated by the set of jump operators appearing in the Lindblad master equation. Using these collective variables, we exactly compute many-body superradiant dynamics for significantly larger system sizes than previously feasible and show explicitly that, unlike in the fully symmetric scenario, the total spin length is generally not conserved throughout the evolution. Furthermore, we analyze the entanglement properties of dark states reached at late times and study their metrological utility by comparing them to symmetric Dicke states. Our methods can be straightforwardly extended to systems with local noise by combining the superspin formalism with SU(4) techniques.

Within the Born-Markov approximation, the evolution of an ensemble of $N$ qubits interacting with an electromagnetic environment is governed by the Lindblad master equation $\dot{\hat{\rho}} = -(i/\hbar) [\hat{H}, \hat{\rho}] + \hat{\mathcal{L}}[\hat{\rho}] $~\cite{Gruner96,Dung02}, where
\begin{equation}\label{lind}
\hat{\mathcal{L}}[\hat{\rho}] =\sum_{i,j=1}^N \frac{\Gamma^{ij}}{2} \biggr( 2\hat{\sigma}_-^j \hat{\rho} \hat{\sigma}_+^i - \{ \hat{\sigma}_+^i\hat{\sigma}_-^j, \hat{\rho}\}\biggr).
\end{equation}
Here, $\hat{H}$ denotes the system Hamiltonian, and $\hat{\sigma}_{\pm}^j$ are the raising and lowering operators of qubit $j$. The coupling rates $\Gamma^{ij}$ encode dissipative interactions among the qubits $i$ and $j$ and are proportional to the imaginary part of the electromagnetic Green's function $\sim \text{Im}\{{\textbf{G}}(\textbf{r}_i,\textbf{r}_j,\omega_0)\}$, evaluated at the qubit resonance frequency $\omega_0$.

Even without coherent Hamiltonian evolution, solving this master equation is hard~\cite{Trivedi2022}, unless the system exhibits a high degree of symmetry. For instance, under full permutational symmetry -- i.e., when $\Gamma^{ij}\equiv \Gamma\; \forall\,i, j$, making the qubits indistinguishable --  the dissipator can be recast in terms of a single collective jump operator $\hat{J}_-=\sum_i \hat{\sigma}_-^i$ that obeys an angular momentum algebra~\cite{Haroche1982}. If the system is initially prepared in a fully symmetric state under qubit exchange, its  evolution remains strictly confined within the $(N+1)$-dimensional subspace spanned by the eigenstates $\{\ket{N/2,m_j}: m_j\in[-N/2,N/2]\}$ of the collective spin operator $\hat{J}_z=\sum_{i}\hat{\sigma}_{z}^i/2$.  This was first described by Dicke in the context of quantum optics~\cite{Dicke54, Haroche1982}, with the resulting fully symmetric eigenstates now known as Dicke states. The efficiency of this description can be rigorously traced back to the Schur-Weyl duality~\cite{GoodmanBook}, which establishes a correspondence between the irreducible representations of the rotation and symmetric groups.

For unstructured 1D baths, the spatially dependent coupling rates appearing in the Lindbladian of Eq.~\eqref{lind} explicitly take the form
\begin{equation}\label{gij}
    \Gamma^{ij} = \ga\cos(kd|i-j|),
\end{equation}
where $\ga$ denotes the decay rate of an individual qubit into the bath, $k\equiv2\pi/\lambda_0=\omega_0/c$ is the wavevector of the guided mode (with $c$ representing the speed of light in the medium), and $d$ is the lattice constant. Such distance-dependent, periodically-modulated couplings naturally appear in various experimental platforms within waveguide quantum electrodynamics (QED), including superconducting qubits coupled to transmission lines \cite{Mirhosseini19,Zanner22}, cold atoms coupled to optical nanofibers~\cite{Vetsch10, Goban12, Gouraud15,Solano17}, quantum dots coupled to nanophotonic waveguides~\cite{Lodahl15, Tiranov23}, and ring resonators~\cite{SI,Hummer2013,Zhou24,Schafer24} supporting degenerate counterpropagating modes.

\begin{figure}[h] 
\begin{center}
\includegraphics[width=8.5cm]{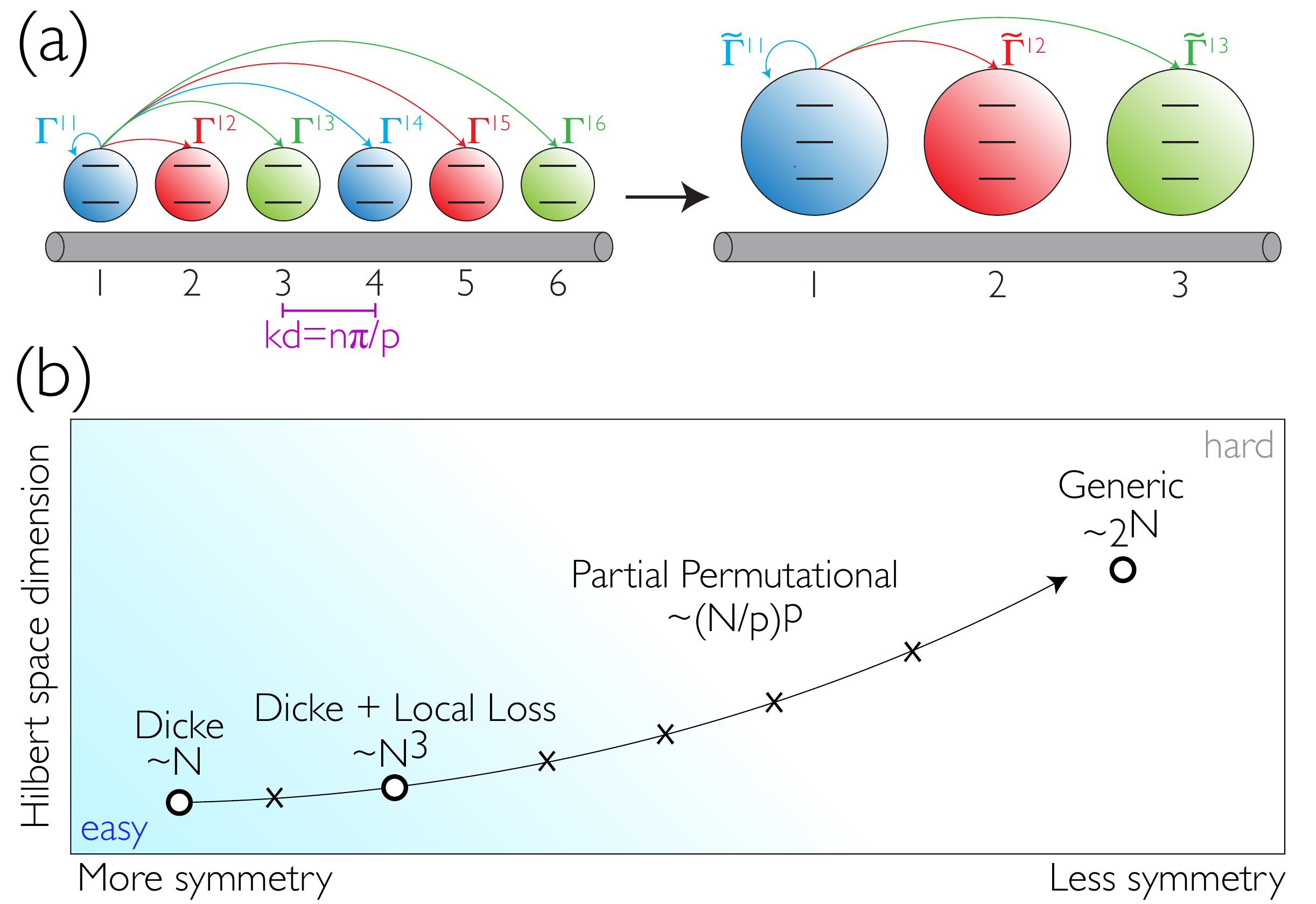}
\caption{\textbf{Exploiting partial permutational symmetry in 1D baths enables efficient and exact computations of quantum dissipative dynamics.} (a)~After integrating out the electromagnetic degrees of freedom, the dissipative couplings $\Gamma^{ij}$ between qubits separated by a distance $kd=n\pi/p$ exhibit partial permutational symmetry if $n, p\in \mathbb{Z}^+$. Under these conditions, the original array of $N$ qubits can be exactly mapped onto $p$ interacting superspins, each possessing $n_a \sim N/p $ levels internal levels. (b)~This mapping constrains the system dynamics to a Hilbert space whose dimension scales polynomially as $\sim(N/p)^p$, thus bridging the gap between ``hard'' problems (without symmetry) and ``easy'' ones (with total permutational symmetry).}
\label{fig:Schematics}
\end{center}
\end{figure}

For lattice constants commensurate with half the resonant wavelength (i.e., satisfying $kd=n\pi/p$, with $p,n\in\mathbb{Z}^+$, or equivalently, $d= n\lambda_0 /2p$), we find that the system dynamics can be described efficiently, exhibiting a computational complexity that scales only polynomially with the number of qubits. This efficiency can be established using two complementary perspectives. The first method involves a direct inspection of the coupling matrix $\boldsymbol{\Gamma}$, whose elements are defined by Eq.~\eqref{gij}, revealing a periodic structure that reflects the underlying symmetry. Alternatively, one can analyze the Lie algebra generated by the jump operators in the master equation, showing explicitly how the symmetry constraints simplify the algebraic structure. We begin with the former approach.

For commensurate lattice constants, the dissipative matrix $\boldsymbol{\Gamma}$ exhibits a periodic structure up to a phase. Specifically, the relation $\Gamma^{i(j+p)} \equiv (-1)^n \Gamma^{ij}$, which follows from Eq.~\eqref{gij} implies that every $p$-th qubit belongs to a permutationally symmetric subset, as shown in Fig.~\ref{fig:Schematics}(a). Consequently, the original $N \times N$ matrix $\boldsymbol{\Gamma}$ can be reduced exactly to a smaller $p\times p$ matrix $\boldsymbol{\tilde{\Gamma}}$, whose elements are directly taken from the first $p\times p$ block of $\boldsymbol{\Gamma}$, i.e., $\tilde{\Gamma}^{ab} = \Gamma^{ab}$ for $a,b \in \{1,2,\dots, p\}$. In terms of these collective degrees of freedom, the dissipator [Eq.~\eqref{lind}] can be compactly written as
\begin{align}
    \hat{\mathcal{L}}[\hat{\rho}] = \sum_{a,b=1}^p \frac{\tilde{\Gamma}^{ab}}{2} \biggr( 2\hat{J}_{b-} \hat{\rho} \hat{J}_{a+} - \{ \hat{J}_{a+}\hat{J}_{b-}, \hat{\rho}\}\biggr),
    \label{superspinEq}
\end{align}
where the collective jump operators are defined by
\begin{equation}\label{eqn:Superspins-Waveguide}
    \hat{J}_{a\pm} = \sum_{l = 0}^{a+lp \leq N} (-1)^{ln} \hat{\sigma}^{a+lp}_\pm,
\end{equation}
with $a \in \{1, 2, \dots p\}$ labeling each of the $p$ collective spins. These collective spin operators satisfy the standard angular momentum commutation relations:
\begin{subequations}\label{eqn:AngMoCommutators}
\begin{align}
    [\hat{J}_{az}, \hat{J}_{a\pm}] &= \pm \hat{J}_{a\pm}, \\
    [\hat{J}_{a+}, \hat{J}_{a-}] &= 2 \hat{J}_{az},
\end{align}
\end{subequations}
with  $\hat{J}_{a z}= \frac{1}{2}\sum_{l = 0}^{a+lp \leq N} \hat{\sigma}^{a+lp}_z$.

Each subset of permutationally symmetric spins can be viewed as a single collective ``superspin'', possessing $n_a+1$ internal levels, where $\sum_{a=1}^p n_a = N$. Concretely, $n_a = \lfloor N/p \rfloor$ for $a>(N \text{~mod~} p)$ and $n_a = \lfloor N/p \rfloor + 1$ otherwise, where $\lfloor x \rfloor$ is the closest integer less than or equal to $x$. As an explicit example, for $N=7$ and $p=3$, we obtain $n_1=3$, and $n_2=n_3=2$. Crucially, superspin operators labeled by different indices commute, as they correspond to disjoint sets of qubits. For each superspin, we  define a collective spin basis $\{\ket{j_a, m_a}\}$ with $j_a \in [0,n_a/2]$ being the total angular momentum and $m_a \in [-j_a, j_a]$. Therefore,
\begin{subequations}
\begin{align}
    \hat{J}_{az}\ket{j_a, m_a} &= m_a\ket{j_a, m_a}, \\
    \hat{J}_{a\pm }\ket{j_a, m_a} &=  \sqrt{(j_a \mp m_a) (j_a \pm m_a + 1)}\ket{j_a, m_a\pm 1}.
\end{align}
\end{subequations} 
For $p=1$ and $p=2$, the superspins reduce to one or two independent Dicke ladders, respectively, whereas intersuperspin interactions are nonzero for any $p>2$.

An alternative approach to directly inspecting the structure of the dissipative matrix $\boldsymbol{\Gamma}$ is to analyze the dimension of the Lie subalgebra generated by the jump operators that appear in the normal form of the master equation, i.e.~\cite{Carmichael00,Clemens03},
\begin{equation}\label{DissipativePart}
\hat{\mathcal{L}}[\hat{\rho}] = \sum_{\mu}\frac{\Gamma_{\mu}}{2} \biggr( 2\hat{\mathcal{O}}_{\mu} \hat{\rho} \hat{\mathcal{O}}_{\mu}^\dagger - \{ \hat{\mathcal{O}}_{\mu}^\dagger\hat{\mathcal{O}}_{\mu}, \hat{\rho}\}\biggr).
\end{equation}
In the above equation $\Gamma_{\mu}$ are collective decay rates (and the eigenvalues of $\boldsymbol{\Gamma}$) and  $\hat{\mathcal{O}}_{\mu} = \sum_{j=1}^N \alpha_{\mu}\hat{\sigma}_-^j$ are collective jump operators, written in terms of the eigenvectors $\alpha_{\mu}$ of $\boldsymbol{\Gamma}$. Different unravelings of the master equation (related by unitary transformations of the jump operators) provide equally valid starting points for describing the dynamics. For instance, in waveguide QED or for spins coupled to ring resonators, a natural unraveling employs directional collective jump operators, which physically correspond to photon emission either to the left or to the right~\cite{Pichler15,Cardenas23}.

For commensurate lattice constants, the collective jump operators (and their adjoints) yield a Lie subalgebra generated by $3p$ elements. More explicitly, this Lie algebra factorizes into a direct sum of $p$ independent $\mathfrak{su}(2)$ algebras, i.e., $\bigoplus_{k=1}^{p}\mathfrak{su}(2)$. Importantly, the finite-dimensional structure emerges irrespective of the chosen unraveling or operator basis. However, identifying the canonical form of the algebra requires additional analysis. In the Supplementary Material (SM)~\cite{SI}, we explicitly show how starting from directional operators it is possible to find the canonical angular momentum representation introduced earlier in Eq.~\eqref{eqn:AngMoCommutators}, reinforcing the direct link between algebraic structure and symmetry properties of the dissipative matrix $\boldsymbol{\Gamma}$.

\begin{figure}[h]
\begin{center}
\includegraphics[width=9cm]{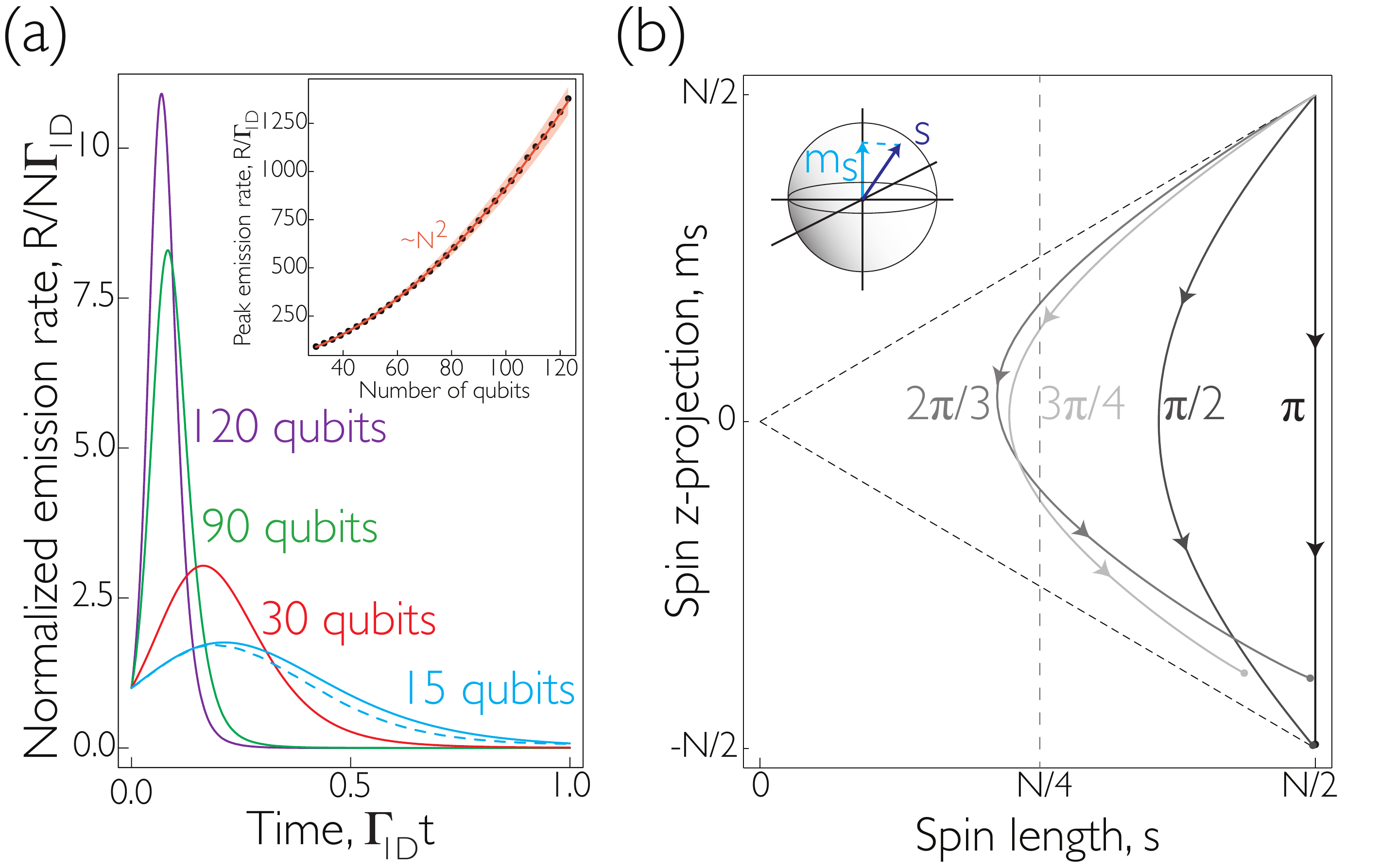}
\caption{\textbf{Exact computation of superradiant decay dynamics for large numbers of qubits.} (a)~Normalized photon emission rate for 15 (blue), 30 (red), 90 (green), and 120 (purple) qubits evolving according to a dissipation-only master equation [with a dissipator in the form of Eqs.~\eqref{lind} and~\eqref{gij}], with lattice constant $kd=2\pi/3$. In waveguide QED, coherent Hamiltonian evolution generally breaks the exact superspin symmetry. Nonetheless, in strongly dissipative regimes such as the early stages of superradiant decay, the superspin formalism remains a reliable approximation, as shown by the close agreement between the dashed and solid blue curves (with and without coherent evolution, respectively). (Inset) Peak emission rate as a function of qubit number for $kd = 2\pi/3$, following the scaling $\sim N^{1.93 \pm 0.01}$, obtained from a power-law fit. Data points are obtained by solving the master equation (up to 81 qubits) and with quantum trajectories (onwards). The shaded region represents the 95\% confidence interval on the fit. (c) Evolution of the spin length $s$ versus its $z$-projection $m_s$ for 36 qubits at different lattice spacings $kd$. Spin length is conserved exclusively at $kd=\pi$ (fully permutational scenario), while for other spacings the spin length notably changes during the dynamics. In both panels, the system is initially prepared in the fully inverted state.}
\label{fig:WGResults}
\end{center}
\end{figure}

Our method significantly reduces the Hilbert space, enabling exact simulations of many-body superradiance in much larger spin ensembles than previously possible, as shown in Fig.~\ref{fig:WGResults}. Transient many-body superradiance was first studied by Dicke~\cite{Dicke54, Haroche1982} in the fully permutationally symmetric limit (i.e., for $p=1$).  In this simplest scenario, the dynamics involve a single collective jump operator -- either fully symmetric or antisymmetric, depending on the parity of $n$ -- and the system evolves due to consecutive quantum jumps between Dicke states starting from a fully inverted state. The more general problem of many-body superradiant decay in spatially extended systems remains open and has recently attracted considerable interest~\cite{Masson22,Sierra22,Robicheaux21,Rubies22,Rubies23,Mok24}.  Although superradiance has been demonstrated to exist in one-dimensional systems, calculations have so far been restricted to exact solutions with only a small number of spins~\cite{Cardenas23}, or to approximate treatments, such as those based on truncated Wigner methods, for larger spin ensembles~\cite{Schneeweiss24, Tebbenjohanns24}. Leveraging the superspin formalism, we are now able to perform exact calculations of superradiant decay involving two jump operators for systems of $\sim 120$ qubits for $p=3$, as shown in Fig.~\ref{fig:WGResults}(a). At around half excitation, we show that the decay rate $R= \sum_{\nu} \Gamma_\nu \braket{\hat{\mathcal{O}}_\nu^\dagger \hat{\mathcal{O}}_\nu }$  scales as $\sim N^2\Gamma_\text{1D}$, resulting in a characteristic photon emission burst.

In the presence of two jump operators, the conventional picture of superradiance in terms of a large collective spin constrained to the surface of the Bloch sphere~\cite{Eberly71, Iemini24} breaks down, as shown in Fig.~\ref{fig:WGResults}(b) for $p>1$. To compute the total spin length, we define the total spin operator as $\hat{\mathbf{S}} = \hat{S}_x\hat{x} + \hat{S}_y\hat{y} + \hat{S}_z\hat{z}$, where
\begin{subequations}
\begin{align}
    \hat{S}_{z} &= \sum_{a=1}^p \hat{J}_{az}, \\
    \hat{S}_{x(y)} &= \frac{1}{2(i)}\sum_{a=1}^p \left(\hat{J}_{a+} \varpm \hat{J}_{a-}\right).
\end{align}
\end{subequations}
As the operators $\hat{J}_{a\pm}$ may carry phases, they are not necessarily equivalent to the standard spin operators $\hat{S}_{x,y} = \frac{1}{2}\sum_{j=1}^N \hat{\sigma}_{x,y}^j$. The average spin length is computed via $\braket{\hat{\mathbf{S}}^2}=s(s+1)$, while its projection along the $z$-axis, $\braket{\hat{S}_z}=m_s$, directly corresponds to the average excitation number in the system. As expected, in the Dicke limit, the total spin length remains maximal throughout the entire decay process, ultimately reaching zero excitations at long times. However, for any other value of $p$, the total spin length is already reduced at the early stages of the decay, with the maximum reduction occurring near half-excitation. This behavior clearly demonstrates that, even with just two jump operators and no coherent Hamiltonian evolution, the physics of many-body decay deviates significantly from the standard paradigm of Dicke superradiance. The collective decay can be represented by a collective spin of maximal length only if the evolution is restricted to a single irreducible representation of SU(2), otherwise $\hat{\mathbf{S}}^2$ is not a Casimir of the algebra (see SM \cite{SI}). Therefore, the spin length is only conserved in the Dicke limit, where all atoms are exchange-symmetric.  
\begin{figure}
\begin{center}
\includegraphics[width=7cm]{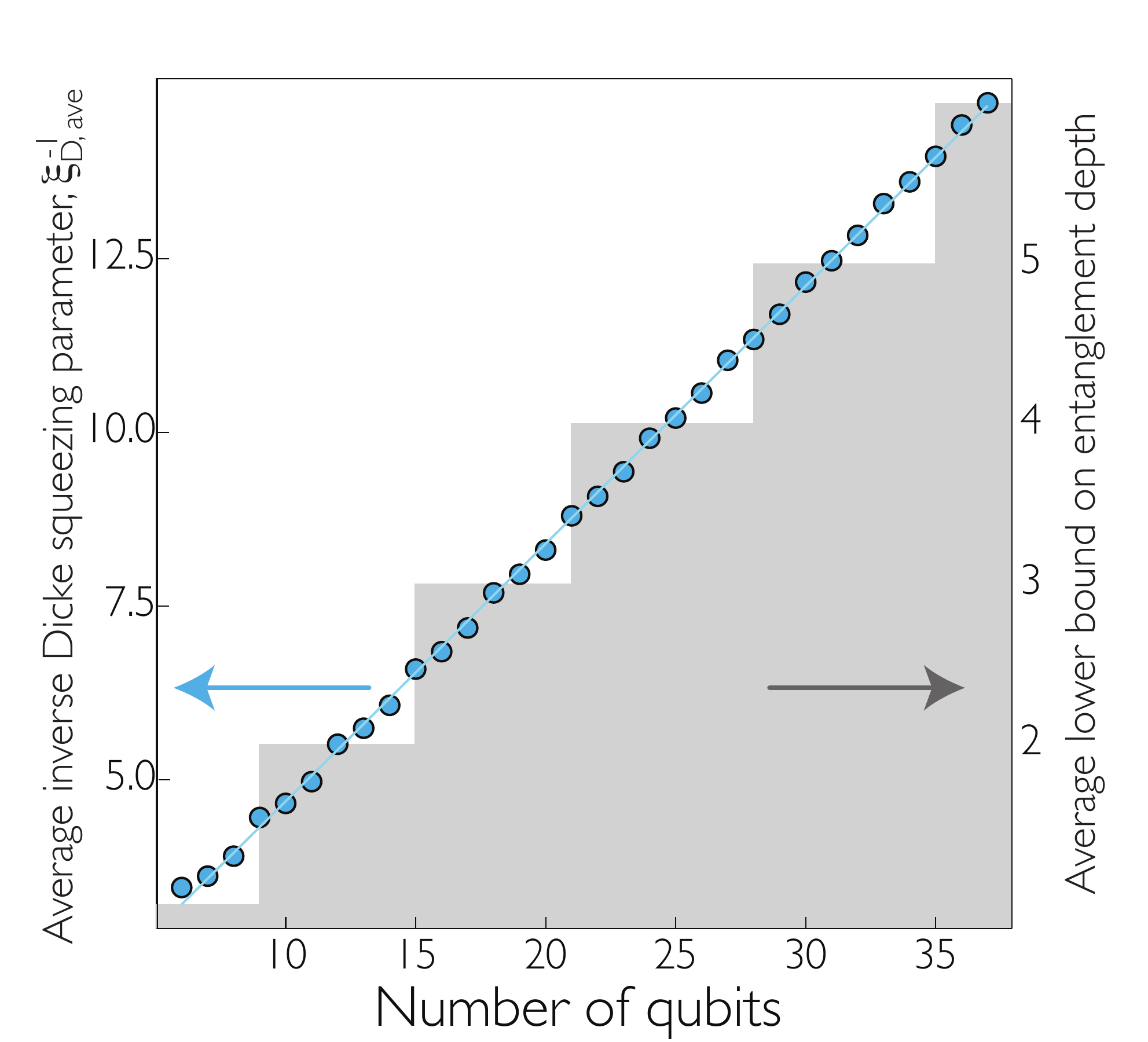}
\caption{\textbf{Entangled dark states for $kd=2\pi/3$ are unique and are metrologically useful.} Average inverse Dicke squeezing parameter as defined in Eq.~\eqref{eqn:DickeAvSqueeze} of the main text (left) and lower bound on entanglement depth (right). The solid blue line shows a linear fit.}
\label{fig:2pi3}
\end{center}
\end{figure}

Another feature of the decay dynamics (for $p\neq 1,2$) is the emergence of dark states at late times. Pure dark states are non-trivial steady states $\ket{\psi}$ such that $\mathcal{L}[\ket{\psi}\bra{\psi}] = 0$. Taking the approach of Ref.~\cite{Kraus08}, we find the dark states by imposing  $\hat{\mathcal{O}}_\nu\ket{\psi} = 0$ (see SM). Remarkably, due to destructive interference, symmetric Dicke states become approximately dark at commensurate lattice spacings when $n N/p$ is even. We find analytically that their decay rates scale as $\sim 1/N^2$ in the low excitation regime (see SM~\cite{SI}). In addition to these nearly dark Dicke states, most configurations also support other dark states within each excitation manifold. This results in a reduced average spin length in the long-time limit.

Solving for dark states in a generic configuration is challenging. In a superspin configuration, however, this task becomes tractable thanks to the reduced number of degrees of freedom. In particular, we focus on the case $kd = 2\pi/3$ with $N$ divisible by 3, where an additional symmetry makes the system especially amenable to analytical treatment. Starting from an initial state, we verify the analytically predicted dark states through numerical evolution. In this configuration, dark states that can be populated via decay from the fully inverted state exist only for at most $1/3$ excitation density. Moreover, each excitation manifold contains a unique such dark state, whose overlap with the corresponding Dicke state approaches unity in the limits of low excitation density and large qubit number (see SM~\cite{SI}). As a result, the maximal spin length is recovered in the long-time limit, as shown in Fig.~\ref{fig:WGResults}(b). 

Decay dynamics starting from a fully inverted initial state for $kd=2\pi/3$ can be exploited to generate metrologically useful dark states~\cite{Burnett93,Klempt11,You18}, with highly excited states being particularly advantageous. Since the dark states are unique in each excitation manifold, photon counting is sufficient to fully identify the specific dark state prepared in a given realization of the decay. Therefore, preparing dark states with a fixed number of excitations beyond the low excitation limit is achievable, although it cannot be done deterministically (see
SM \cite{SI}). 

Harnessing partial permutational symmetry enables us to efficiently characterize the entanglement properties of the generated dark states. To quantify entanglement, we employ the Dicke squeezing parameter~\cite{Duan11, Duan14}, defined as
\begin{equation}
    \label{eqn:DickeSqueeze}
    \xi_D = N\frac{(\Delta \hat{S}_z)^2 + 1/4}{\braket{\hat{S}_x^2 + \hat{S}_y^2}}.
\end{equation}
The Dicke squeezing parameter imposes a rigorous lower bound on the entanglement depth as $\lceil\xi_D^{-1} \rceil - 2$, where $\lceil x \rceil$ is the closest integer greater than or equal to $x$~\cite{Duan11, Duan14}. States with $\xi_D<0.5$ are entangled and exhibit enhanced phase sensitivity over that of uncorrelated qubits. Denoting as $p_m$ the probability of ending up in the dark state with $m$ excitations, the average inverse Dicke squeezing parameter (conditioned to the measurement of $N-m$ photons for $m\in [1, \dots, N/3]$) is
\begin{equation}
\label{eqn:DickeAvSqueeze}
    \xi_{D, \text{ave}}^{-1} = \sum_{m=1}^{N/3} p_m\xi_{D, m}^{-1},
\end{equation}
where $\xi_{D, m}$ is the Dicke squeezing parameter of the state in the manifold $m$. Figure~\ref{fig:2pi3} shows the linear increase of $\xi_{D, \text{ave}}^{-1}$ with qubit number, approaching Heisenberg scaling.

In summary, we have demonstrated that partial permutational symmetry (which is characteristic of dissipators of 1D baths) can be exploited to exactly solve quantum many-body dynamics with only polynomial computational complexity. Precise control over the positions of individual emitters allows tuning the degree of symmetry in the system, with the potential to drastically alter its collective optical properties. Ideal candidates to explore this physics are cold atoms near microring resonators~\cite{Zhou24,Suresh25,Zhou25} or in ring and bow-tie cavities~\cite{Schafer24, Simon20, Chen22}, superconducting qubits coupled to transmission lines~\cite{Mirhosseini19}, and color centers coupled to microresonators~\cite{Lukin23}.  Another promising avenue for the application of the superspin formalism is in describing qubits coupled to multimode cavities~\cite{Kollar15, Vaidya18}, where a partially permutational symmetric basis emerges as a natural choice for ensembles at different spatial points~\cite{Marsh24}. 

We note, however, that the validity of the superspin picture relies on the Hamiltonian evolution respecting the same partial permutational symmetry as the dissipator. Systems of qubits resonantly coupled to ring resonators naturally preserve such symmetry~\cite{Lukin25}. In contrast, qubits coupled to open-ended waveguides undergo a Hamiltonian evolution given by
\begin{align}
\hat{H}= \sum_{i,j=1}^N \frac{\Gamma_{1\text{D}}}{2}\sin(kd|i-j|) \hat{\sigma}_+^i \hat{\sigma}_-^j, 
\end{align}
which cannot be written in terms of superspins.  Therefore, the superspin representation remains a good approximation primarily in regimes where dissipative dynamics dominate and coherent evolution can be neglected, such as during rapid superradiant bursts. Alternatively, experimental schemes could be engineered to effectively nullify coherent interactions. For instance, one could simultaneously couple emitters to two waveguides with respective wavevectors $k_i$ and spacings  $d_i$ fulfilling $k_2d_2 = 2\pi l - k_1d_1$ such that $\hat{H}(k_1d_1) + \hat{H}(k_2d_2) =0$, or employ dynamical decoupling methods to average out Hamiltonian terms~\cite{Waugh68}. Another reason for the breaking of the partial permutational symmetry is disorder. Numerical results (see SM~\cite{SI}) for small qubit numbers indicate that relevant observables are generally robust to small levels of noise. These findings further support the practical use of partial permutational symmetry as a coarse-graining approximation, with improved accuracy achievable by increasing the number of superspins at the expense of computational resources.

Finally, superspins provide an exact benchmark for validating approximate techniques for open quantum dynamics, such as cumulants~\cite{Robicheaux21_2,Plankensteiner22,Rubies23,Holzinger24}, truncated Wigner methods~\cite{Mink23,Tebbenjohanns24,Guardiola25}, and tensor networks \cite{Verstraete04, Manzoni17, Malz2024}. Related ideas involving partial permutational symmetry have also been used to benchmark approximate simulations of Hamiltonian dynamics -- for example, in two-dimensional trapped-ion systems, where the crystal is modeled as a set of concentric rings and angular momentum operators are defined for each ring~\cite{Shankar22}. Our formalism can be straightforwardly extended to incorporate symmetry-breaking processes at the single-qubit level (e.g., local spontaneous emission into additional baths), leveraging previously developed SU(4) techniques that scale at worst as $(\sim N^{3p})$ \cite{Holland13, SI}. Furthermore, it would be intriguing to extend existing phase-space methods that exploit this symmetry \cite{Forbes24}. Partial permutational symmetry should also be useful for the efficient description of multilevel atoms or qudits~\cite{Bolanos15}, expanding the applicability of our approach to even broader classes of quantum systems.

\textit{Acknowledgments --} We are thankful to P. Barberis-Blostein, D. Chang, M.J. Holland,  D. Malz, W.-K. Mok, and D. Wild  for stimulating discussions and suggestions. We acknowledge support by the National Science Foundation through the CAREER Award (No. 2047380), the Air Force Office of Scientific Research through their Young Investigator Prize (grant No. 21RT0751), as well as by the David and Lucile Packard Foundation. J. T. Lee acknowledges support by the National Science Foundation Graduate Research Fellowship under grant No. DGE 2036197. This research was supported in part by grant NSF PHY-2309135 to the Kavli Institute for Theoretical Physics (KITP).

\bibliography{mainref}

@article{Dicke54, 
  title = {{Coherence in Spontaneous Radiation Processes}},
  author = {Dicke, R. H.},
  journal = {Phys. Rev.},
  volume = {93},
  issue = {1},
  pages = {99--110},
  numpages = {0},
  year = {1954},
  month = {Jan},
  publisher = {American Physical Society},
  doi = {10.1103/PhysRev.93.99},
  url = {https://link.aps.org/doi/10.1103/PhysRev.93.99}
}

@Article{Masson22,
author={Masson, Stuart J.
and Asenjo-Garcia, Ana},
title={{Universality of Dicke superradiance in arrays of quantum emitters}},
journal={Nature Communications},
year={2022},
month={Apr},
day={27},
volume={13},
number={1},
pages={2285},
issn={2041-1723},
doi={10.1038/s41467-022-29805-4},
url={https://doi.org/10.1038/s41467-022-29805-4}
}

@article{Sierra22,
  title = {{Dicke Superradiance in Ordered Lattices: Dimensionality Matters}},
  author = {Sierra, Eric and Masson, Stuart J. and Asenjo-Garcia, Ana},
  journal = {Phys. Rev. Res.},
  volume = {4},
  issue = {2},
  pages = {023207},
  numpages = {11},
  year = {2022},
  month = {Jun},
  publisher = {American Physical Society},
  doi = {10.1103/PhysRevResearch.4.023207},
  url = {https://link.aps.org/doi/10.1103/PhysRevResearch.4.023207}
}

@article{Robicheaux21,
  title = {{Theoretical study of early-time superradiance for atom clouds and arrays}},
  author = {Robicheaux, F.},
  journal = {Phys. Rev. A},
  volume = {104},
  issue = {6},
  pages = {063706},
  numpages = {10},
  year = {2021},
  month = {Dec},
  publisher = {American Physical Society},
  doi = {10.1103/PhysRevA.104.063706},
  url = {https://link.aps.org/doi/10.1103/PhysRevA.104.063706}
}

@misc{Mok24,
      title={{Universal scaling laws for correlated decay of many-body quantum systems}}, 
      author={Wai-Keong Mok and Avishi Poddar and Eric Sierra and Cosimo C. Rusconi and John Preskill and Ana Asenjo-Garcia},
      year={2024},
      archivePrefix={arXiv},
      eprint={2406.00722},
      primaryClass={quant-ph},
      url={https://arxiv.org/abs/2406.00722}, 
}

@article{Robicheaux21_2,
  title = {{Beyond lowest order mean-field theory for light interacting with atom arrays}},
  author = {Robicheaux, F. and Suresh, Deepak A.},
  journal = {Phys. Rev. A},
  volume = {104},
  issue = {2},
  pages = {023702},
  numpages = {12},
  year = {2021},
  month = {Aug},
  publisher = {American Physical Society},
  doi = {10.1103/PhysRevA.104.023702},
  url = {https://link.aps.org/doi/10.1103/PhysRevA.104.023702}
}

@article{Rubies23,
  title = {{Characterizing superradiant dynamics in atomic arrays via a cumulant expansion approach}},
  author = {Rubies-Bigorda, Oriol and Ostermann, Stefan and Yelin, Susanne F.},
  journal = {Phys. Rev. Res.},
  volume = {5},
  issue = {1},
  pages = {013091},
  numpages = {12},
  year = {2023},
  month = {Feb},
  publisher = {American Physical Society},
  doi = {10.1103/PhysRevResearch.5.013091},
  url = {https://link.aps.org/doi/10.1103/PhysRevResearch.5.013091}
}

@Article{Mink23,
	title={{Collective radiative interactions in the discrete truncated Wigner approximation}},
	author={Christopher D. Mink and Michael Fleischhauer},
	journal={SciPost Phys.},
	volume={15},
	pages={233},
	year={2023},
	publisher={SciPost},
	doi={10.21468/SciPostPhys.15.6.233},
	url={https://scipost.org/10.21468/SciPostPhys.15.6.233},
}

@article{Rubies22,
  title = {{Superradiance and subradiance in inverted atomic arrays}},
  author = {Rubies-Bigorda, Oriol and Yelin, Susanne F.},
  journal = {Phys. Rev. A},
  volume = {106},
  issue = {5},
  pages = {053717},
  numpages = {12},
  year = {2022},
  month = {Nov},
  publisher = {American Physical Society},
  doi = {10.1103/PhysRevA.106.053717},
  url = {https://link.aps.org/doi/10.1103/PhysRevA.106.053717}
}

@article{Cardenas23,
  title = {{Many-Body Superradiance and Dynamical Mirror Symmetry Breaking in Waveguide QED}},
  author = {Cardenas-Lopez, Silvia and Masson, Stuart J. and Zager, Zoe and Asenjo-Garcia, Ana},
  journal = {Phys. Rev. Lett.},
  volume = {131},
  issue = {3},
  pages = {033605},
  numpages = {7},
  year = {2023},
  month = {Jul},
  publisher = {American Physical Society},
  doi = {10.1103/PhysRevLett.131.033605},
  url = {https://link.aps.org/doi/10.1103/PhysRevLett.131.033605}
}

@article{Molmer18, 
  title = {{Monte-Carlo simulations of superradiant lasing}},
  author = {Zhang, Y. and Zhang, Y. and Mølmer, K.},
  journal = {New J. Phys.},
  volume = {20},
  issue = {},
  pages = {112001},
  numpages = {0},
  year = {2018},
  month = {},
  publisher = {IOP},
  doi = {10.1088/1367-2630/aaec36},
  url = {https://iopscience.iop.org/article/10.1088/1367-2630/aaec36}
}

@article{Duan14, 
  title = {{Quantum metrology with Dicke squeezed states}},
  author = {Zhang, Z. and Duan, L. M.},
  journal = {New J. Phys.},
  volume = {16},
  issue = {},
  pages = {103037},
  numpages = {0},
  year = {2014},
  month = {},
  publisher = {IOP},
  doi = {10.1088/1367-2630/16/10/103037},
  url = {https://iopscience.iop.org/article/10.1088/1367-2630/16/10/103037}
}

@article{Sarkar87,
doi = {10.1088/0305-4470/20/8/028},
url = {https://dx.doi.org/10.1088/0305-4470/20/8/028},
year = {1987},
month = {jun},
publisher = {},
volume = {20},
number = {8},
pages = {2147},
author = {S Sarkar and  J S Satchell},
title = {{Solution of master equations for small bistable systems}},
journal = {Journal of Physics A: Mathematical and General}
}

@article{Chase08,
  title = {{Collective processes of an ensemble of spin-$1/2$ particles}},
  author = {Chase, Bradley A. and Geremia, J. M.},
  journal = {Phys. Rev. A},
  volume = {78},
  issue = {5},
  pages = {052101},
  numpages = {14},
  year = {2008},
  month = {Nov},
  publisher = {American Physical Society},
  doi = {10.1103/PhysRevA.78.052101},
  url = {https://link.aps.org/doi/10.1103/PhysRevA.78.052101}
}

@article{Holland13,
  title = {{Simulating open quantum systems by applying SU(4) to quantum master equations}},
  author = {Xu, Minghui and Tieri, D. A. and Holland, M. J.},
  journal = {Phys. Rev. A},
  volume = {87},
  issue = {6},
  pages = {062101},
  numpages = {7},
  year = {2013},
  month = {Jun},
  publisher = {American Physical Society},
  doi = {10.1103/PhysRevA.87.062101},
  url = {https://link.aps.org/doi/10.1103/PhysRevA.87.062101}
}

@article{Hartmann16,
author = {Hartmann, Stephan},
title = {{Generalized Dicke states}},
year = {2016},
issue_date = {November 2016},
publisher = {Rinton Press, Incorporated},
address = {Paramus, NJ},
volume = {16},
number = {15–16},
issn = {1533-7146},
journal = {Quantum Info. Comput.},
month = {nov},
pages = {1333–1348},
numpages = {16},
}

@article{Zhou24,
  title = {Trapped Atoms and Superradiance on an Integrated Nanophotonic Microring Circuit},
  author = {Zhou, Xinchao and Tamura, Hikaru and Chang, Tzu-Han and Hung, Chen-Lung},
  journal = {Phys. Rev. X},
  volume = {14},
  issue = {3},
  pages = {031004},
  numpages = {11},
  year = {2024},
  month = {Jul},
  publisher = {American Physical Society},
  doi = {10.1103/PhysRevX.14.031004},
  url = {https://link.aps.org/doi/10.1103/PhysRevX.14.031004}
}

@article{Hummer2013, 
  title = {{Weak and strong coupling regimes in plasmonic QED}},
  author = {H\"ummer, T. and Garc\'{\i}a-Vidal, F. J. and Mart\'{\i}n-Moreno, L. and Zueco, D.},
  journal = {Phys. Rev. B},
  volume = {87},
  issue = {11},
  pages = {115419},
  numpages = {16},
  year = {2013},
  month = {Mar},
  publisher = {American Physical Society},
  doi = {10.1103/PhysRevB.87.115419},
  url = {https://link.aps.org/doi/10.1103/PhysRevB.87.115419}
}

@article{Gruner96,
  title = {{Green-function approach to the radiation-field quantization for homogeneous and inhomogeneous Kramers-Kronig dielectrics}},
  author = {Gruner, T. and Welsch, D.-G.},
  journal = {Phys. Rev. A},
  volume = {53},
  issue = {3},
  pages = {1818--1829},
  numpages = {0},
  year = {1996},
  month = {Mar},
  publisher = {American Physical Society},
  doi = {10.1103/PhysRevA.53.1818},
  url = {https://link.aps.org/doi/10.1103/PhysRevA.53.1818}
}

@article{Dung02,
  title = {{Resonant dipole-dipole interaction in the presence of dispersing and absorbing surroundings}},
  author = {Dung, Ho Trung and Kn\"oll, Ludwig and Welsch, Dirk-Gunnar},
  journal = {Phys. Rev. A},
  volume = {66},
  issue = {6},
  pages = {063810},
  numpages = {16},
  year = {2002},
  month = {Dec},
  publisher = {American Physical Society},
  doi = {10.1103/PhysRevA.66.063810},
  url = {https://link.aps.org/doi/10.1103/PhysRevA.66.063810}
}

@article{Pichler15,
  title = {{Quantum optics of chiral spin networks}},
  author = {Pichler, Hannes and Ramos, Tom\'as and Daley, Andrew J. and Zoller, Peter},
  journal = {Phys. Rev. A},
  volume = {91},
  issue = {4},
  pages = {042116},
  numpages = {19},
  year = {2015},
  month = {Apr},
  publisher = {American Physical Society},
  doi = {10.1103/PhysRevA.91.042116},
  url = {https://link.aps.org/doi/10.1103/PhysRevA.91.042116}
}

@article{Trivedi2022,
  title = {{Transitions in Computational Complexity of Continuous-Time Local Open Quantum Dynamics}},
  author = {Trivedi, Rahul and Cirac, J. Ignacio},
  journal = {Phys. Rev. Lett.},
  volume = {129},
  issue = {26},
  pages = {260405},
  numpages = {7},
  year = {2022},
  month = {Dec},
  publisher = {American Physical Society},
  doi = {10.1103/PhysRevLett.129.260405},
  url = {https://link.aps.org/doi/10.1103/PhysRevLett.129.260405}
}

@article{Haroche1982,
title = {{Superradiance: An essay on the theory of collective spontaneous emission}},
journal = {Physics Reports},
volume = {93},
number = {5},
pages = {301-396},
year = {1982},
issn = {0370-1573},
doi = {https://doi.org/10.1016/0370-1573(82)90102-8},
url = {https://www.sciencedirect.com/science/article/pii/0370157382901028},
author = {M. Gross and S. Haroche},
}

@article{Iemini24,
  title = {Dynamics of inhomogeneous spin ensembles with all-to-all interactions: Breaking permutational invariance},
  author = {Iemini, Fernando and Chang, Darrick and Marino, Jamir},
  journal = {Phys. Rev. A},
  volume = {109},
  issue = {3},
  pages = {032204},
  numpages = {11},
  year = {2024},
  month = {Mar},
  publisher = {American Physical Society},
  doi = {10.1103/PhysRevA.109.032204},
  url = {https://link.aps.org/doi/10.1103/PhysRevA.109.032204}
}

@article{Clemens03,
  title = {Collective spontaneous emission from a line of atoms},
  author = {Clemens, J. P. and Horvath, L. and Sanders, B. C. and Carmichael, H. J.},
  journal = {Phys. Rev. A},
  volume = {68},
  issue = {2},
  pages = {023809},
  numpages = {19},
  year = {2003},
  month = {Aug},
  publisher = {American Physical Society},
  doi = {10.1103/PhysRevA.68.023809},
  url = {https://link.aps.org/doi/10.1103/PhysRevA.68.023809}
}

@article{Forbes24,
author = {Andrew Kolmer Forbes and Philip Daniel Blocher and Ivan H. Deutsch},
journal = {Optica Quantum},
number = {5},
pages = {310--328},
publisher = {Optica Publishing Group},
title = {Modeling local decoherence of a spin ensemble using a generalized Holstein-Primakoff mapping to a bosonic mode},
volume = {2},
month = {Oct},
year = {2024},
url = {https://opg.optica.org/opticaq/abstract.cfm?URI=opticaq-2-5-310},
doi = {10.1364/OPTICAQ.528078},
}

@article{AnaPRA,
  title = {{Atom-light interactions in quasi-one-dimensional nanostructures: A Green's-function perspective}},
  author = {Asenjo-Garcia, A. and Hood, J. D. and Chang, D. E. and Kimble, H. J.},
  journal = {Phys. Rev. A},
  volume = {95},
  issue = {3},
  pages = {033818},
  numpages = {16},
  year = {2017},
  month = {Mar},
  publisher = {American Physical Society},
  doi = {10.1103/PhysRevA.95.033818},
  url = {https://link.aps.org/doi/10.1103/PhysRevA.95.033818}
}

@article{Burnett93,
  title = {{Interferometric detection of optical phase shifts at the Heisenberg limit}},
  author = {Holland, M. J. and Burnett, K.},
  journal = {Phys. Rev. Lett.},
  volume = {71},
  issue = {9},
  pages = {1355--1358},
  numpages = {0},
  year = {1993},
  month = {Aug},
  publisher = {American Physical Society},
  doi = {10.1103/PhysRevLett.71.1355},
  url = {https://link.aps.org/doi/10.1103/PhysRevLett.71.1355}
}

@article{Klempt11,
author = {B. Lücke  and M. Scherer  and J. Kruse  and L. Pezzé  and F. Deuretzbacher  and P. Hyllus  and O. Topic  and J. Peise  and W. Ertmer  and J. Arlt  and L. Santos  and A. Smerzi  and C. Klempt },
title = {{Twin Matter Waves for Interferometry Beyond the Classical Limit}},
journal = {Science},
volume = {334},
number = {6057},
pages = {773-776},
year = {2011},
doi = {10.1126/science.1208798},
URL = {https://www.science.org/doi/abs/10.1126/science.1208798}}

@article{You18,
author = {Yi-Quan Zou  and Ling-Na Wu  and Qi Liu  and Xin-Yu Luo  and Shuai-Feng Guo  and Jia-Hao Cao  and Meng Khoon Tey  and Li You },
title = {{Beating the classical precision limit with spin-1 Dicke states of more than 10,000 atoms}},
journal = {Proceedings of the National Academy of Sciences},
volume = {115},
number = {25},
pages = {6381-6385},
year = {2018},
doi = {10.1073/pnas.1715105115},
URL = {https://www.pnas.org/doi/abs/10.1073/pnas.1715105115}}

@article{Schneeweiss24,
  title = {{Observation of Superradiant Bursts in a Cascaded Quantum System}},
  author = {Liedl, Christian and Tebbenjohanns, Felix and Bach, Constanze and Pucher, Sebastian and Rauschenbeutel, Arno and Schneeweiss, Philipp},
  journal = {Phys. Rev. X},
  volume = {14},
  issue = {1},
  pages = {011020},
  numpages = {15},
  year = {2024},
  month = {Feb},
  publisher = {American Physical Society},
  doi = {10.1103/PhysRevX.14.011020},
  url = {https://link.aps.org/doi/10.1103/PhysRevX.14.011020}
}

@article{Duan11,
  author = {Duan, L.-M.},
  title = {{Entanglement Detection in the Vicinity of Arbitrary Dicke States}},
  journal = {Phys. Rev. Lett.},
  volume = {107},
  issue = {18},
  pages = {180502},
  numpages = {4},
  year = {2011},
  month = {Oct},
  publisher = {American Physical Society},
  doi = {10.1103/PhysRevLett.107.180502},
  url = {https://link.aps.org/doi/10.1103/PhysRevLett.107.180502}
}

@article{Marsh24,
  title = {{Entanglement and Replica Symmetry Breaking in a Driven-Dissipative Quantum Spin Glass}},
  author = {Marsh, Brendan P. and Kroeze, Ronen M. and Ganguli, Surya and Gopalakrishnan, Sarang and Keeling, Jonathan and Lev, Benjamin L.},
  journal = {Phys. Rev. X},
  volume = {14},
  issue = {1},
  pages = {011026},
  numpages = {24},
  year = {2024},
  month = {Feb},
  publisher = {American Physical Society},
  doi = {10.1103/PhysRevX.14.011026},
  url = {https://link.aps.org/doi/10.1103/PhysRevX.14.011026}
}

@article{Vaidya18,
  title = {{Tunable-Range, Photon-Mediated Atomic Interactions in Multimode Cavity QED}},
  author = {Vaidya, Varun D. and Guo, Yudan and Kroeze, Ronen M. and Ballantine, Kyle E. and Koll\'ar, Alicia J. and Keeling, Jonathan and Lev, Benjamin L.},
  journal = {Phys. Rev. X},
  volume = {8},
  issue = {1},
  pages = {011002},
  numpages = {17},
  year = {2018},
  month = {Jan},
  publisher = {American Physical Society},
  doi = {10.1103/PhysRevX.8.011002},
  url = {https://link.aps.org/doi/10.1103/PhysRevX.8.011002}
}

@article{Kollar15, 
  title = {{An adjustable-length cavity and Bose–Einstein condensate apparatus for multimode cavity QED}},
  author = {Kollár, Alicia J. and Papageorge, Alexander T. and Baumann, Kristian and Armen, Michael A. and Lev, Benjamin L.},
  journal = {New J. Phys.},
  volume = {17},
  issue = {},
  pages = {043012},
  numpages = {0},
  year = {2015},
  month = {},
  publisher = {IOP},
  doi = {10.1088/1367-2630/17/4/043012},
  url = {https://iopscience.iop.org/article/10.1088/1367-2630/17/4/043012/meta}
}

@book{Tai,
  title     = "Dyadic Green's functions in electromagnetic theory",
  author    = "Tai, Chen-to",
  year      = 1994,
  publisher = "IEEE Press",
  note      = {}
}

@article{Simon20,
  title = {{Observation of Laughlin states made of light}},
  author = {Clark, Logan W. and Schine, Nathan and Baum, Claire and Jia, Ningyuan and Simon, Jonathan},
  journal = {Nature},
  volume = {582},
  issue = {26},
  pages = {41-45}, 
  numpages = {5},
  year = {2020},
  month = {June},
  publisher = {Nature},
  doi = {10.1038/s41586-020-2318-5},
  url = {https://www.nature.com/articles/s41586-020-2318-5}
}

@article{Chen22,
author = {Yu-Ting Chen and Michal Szurek and Beili Hu and Julius de Hond and Boris Braverman and Vladan Vuletic},
journal = {Opt. Express},
number = {21},
pages = {37426--37435},
publisher = {Optica Publishing Group},
title = {{High finesse bow-tie cavity for strong atom-photon coupling in Rydberg arrays}},
volume = {30},
month = {Oct},
year = {2022},
url = {https://opg.optica.org/oe/abstract.cfm?URI=oe-30-21-37426},
doi = {10.1364/OE.469644},
}

@article{Cooper24,
	author = {Cooper, Eric S. and Kunkel, Philipp and Periwal, Avikar and Schleier-Smith, Monika},
	date = {2024/05/01},
	doi = {10.1038/s41567-024-02407-1},
	id = {Cooper2024},
	isbn = {1745-2481},
	journal = {Nat. Phys.},
	number = {5},
	pages = {770--775},
	title = {Graph states of atomic ensembles engineered by photon-mediated entanglement},
	volume = {20},
	year = {2024}}

@article{Colombo22,
	author = {Colombo, Simone and Pedrozo-Pe{\~n}afiel, Edwin and Adiyatullin, Albert F. and Li, Zeyang and Mendez, Enrique and Shu, Chi and Vuleti{\'c}, Vladan},
	date = {2022/08/01},
	doi = {10.1038/s41567-022-01653-5},
	id = {Colombo2022},
	isbn = {1745-2481},
	journal = {Nat. Phys.},
	number = {8},
	pages = {925--930},
	title = {Time-reversal-based quantum metrology with many-body entangled states},
	url = {https://doi.org/10.1038/s41567-022-01653-5},
	volume = {18},
	year = {2022}}

@article{Cox16,
	author = {Cox, Kevin C. and Greve, Graham P. and Weiner, Joshua M. and Thompson, James K.},
	doi = {10.1103/PhysRevLett.116.093602},
	issue = {9},
	journal = {Phys. Rev. Lett.},
	month = {Mar},
	numpages = {5},
	pages = {093602},
	publisher = {American Physical Society},
	title = {Deterministic Squeezed States with Collective Measurements and Feedback},
	url = {https://link.aps.org/doi/10.1103/PhysRevLett.116.093602},
	volume = {116},
	year = {2016}}

@article{Hosten16,
	author = {Hosten, Onur and Engelsen, Nils J. and Krishnakumar, Rajiv and Kasevich, Mark A.},
	date = {2016/01/28/print},
	day = {28},
	isbn = {0028-0836},
	journal = {Nature},
	l3 = {10.1038/nature16176},
	m3 = {Letter},
	month = {01},
	number = {7587},
	pages = {505--508},
	publisher = {Nature Publishing Group, a division of Macmillan Publishers Limited. All Rights Reserved.},
	title = {Measurement noise 100 times lower than the quantum-projection limit using entangled atoms},
	ty = {JOUR},
	url = {http://dx.doi.org/10.1038/nature16176},
	volume = {529},
	year = {2016}}

@article{Mivehvar21,
	author = {Mivehvar, Farokh and Piazza, Francesco and Donner, Tobias and Ritsch, Helmut},
	date = {2021/01/02},
	doi = {10.1080/00018732.2021.1969727},
	isbn = {0001-8732},
	journal = {Adv. Phys.},
	journal1 = {Advances in Physics},
	journal2 = {Advances in Physics},
	month = {01},
	number = {1},
	pages = {1--153},
	publisher = {Taylor \& Francis},
	title = {Cavity {QED} with quantum gases: new paradigms in many-body physics},
	type = {doi: 10.1080/00018732.2021.1969727},
	url = {https://doi.org/10.1080/00018732.2021.1969727},
	volume = {70},
	year = {2021},
	year1 = {2021}}

@article{Ritsch13,
	author = {Ritsch, Helmut and Domokos, Peter and Brennecke, Ferdinand and Esslinger, Tilman},
	doi = {10.1103/RevModPhys.85.553},
	issue = {2},
	journal = {Rev. Mod. Phys.},
	month = {Apr},
	numpages = {0},
	pages = {553--601},
	publisher = {American Physical Society},
	title = {Cold atoms in cavity-generated dynamical optical potentials},
	url = {https://link.aps.org/doi/10.1103/RevModPhys.85.553},
	volume = {85},
	year = {2013}}

@article{Baumann10,
	author = {Baumann, Kristian and Guerlin, Christine and Brennecke, Ferdinand and Esslinger, Tilman},
	date = {2010/04/01},
	doi = {10.1038/nature09009},
	id = {Baumann2010},
	isbn = {1476-4687},
	journal = {Nature},
	number = {7293},
	pages = {1301--1306},
	title = {Dicke quantum phase transition with a superfluid gas in an optical cavity},
	url = {https://doi.org/10.1038/nature09009},
	volume = {464},
	year = {2010}}

@article{Young24,
	author = {Young, Dylan J. and Chu, Anjun and Song, Eric Yilun and Barberena, Diego and Wellnitz, David and Niu, Zhijing and Sch\"{a}fer, Vera M. and Lewis-Swan, Robert J. and Rey, Ana Maria and Thompson, James K.},
	date = {2024/01/01},
	doi = {10.1038/s41586-023-06911-x},
	id = {Young2024},
	isbn = {1476-4687},
	journal = {Nature},
	number = {7996},
	pages = {679--684},
	title = {Observing dynamical phases of {BCS} superconductors in a cavity {QED} simulator},
	url = {https://doi.org/10.1038/s41586-023-06911-x},
	volume = {625},
	year = {2024}}

@article{Periwal21,
	author = {Periwal, Avikar and Cooper, Eric S. and Kunkel, Philipp and Wienand, Julian F. and Davis, Emily J. and Schleier-Smith, Monika},
	date = {2021/12/01},
	doi = {10.1038/s41586-021-04156-0},
	id = {Periwal2021},
	isbn = {1476-4687},
	journal = {Nature},
	number = {7890},
	pages = {630--635},
	title = {Programmable interactions and emergent geometry in an array of atom clouds},
	url = {https://doi.org/10.1038/s41586-021-04156-0},
	volume = {600},
	year = {2021}}

@article{Bolanos15,
	author = {Bola\~{n}os, Marduk and Barberis-Blostein, Pablo},
	date = {2015/10/08},
	doi = {10.1088/1751-8113/48/44/445301},
	isbn = {1751-8121; 1751-8113},
	journal = {J. Phys. A},
	number = {44},
	pages = {445301},
	publisher = {IOP Publishing},
	title = {Algebraic solution of the {L}indblad equation for a collection of multilevel systems coupled to independent environments},
	url = {https://dx.doi.org/10.1088/1751-8113/48/44/445301},
	volume = {48},
	year = {2015}}

@article{Bohnet12,
	author = {Bohnet, Justin G. and Chen, Zilong and Weiner, Joshua M. and Meiser, Dominic and Holland, Murray J. and Thompson, James K.},
	da = {2012/04/01},
	doi = {10.1038/nature10920},
	id = {Bohnet2012},
	isbn = {1476-4687},
	journal = {Nature},
	number = {7392},
	pages = {78--81},
	title = {A steady-state superradiant laser with less than one intracavity photon},
	ty = {JOUR},
	url = {https://doi.org/10.1038/nature10920},
	volume = {484},
	year = {2012}}

@article{Alvarez20,
	author = {Alvarez-Giron, W and Barberis-Blostein, P},
	date = {2020/10/06},
	doi = {10.1088/1751-8121/abb1e2},
	isbn = {1751-8121; 1751-8113},
	journal = {J. Phys. A},
	number = {43},
	pages = {435301},
	publisher = {IOP Publishing},
	title = {The atomic damping basis and the collective decay of interacting two-level atoms},
	url = {https://dx.doi.org/10.1088/1751-8121/abb1e2},
	volume = {53},
	year = {2020}}

@article{Meiser09,
  title = {Prospects for a Millihertz-Linewidth Laser},
  author = {Meiser, D. and Ye, Jun and Carlson, D. R. and Holland, M. J.},
  journal = {Phys. Rev. Lett.},
  volume = {102},
  issue = {16},
  pages = {163601},
  numpages = {4},
  year = {2009},
  month = {Apr},
  publisher = {American Physical Society},
  doi = {10.1103/PhysRevLett.102.163601},
  url = {https://link.aps.org/doi/10.1103/PhysRevLett.102.163601}
}

@article{Tebbenjohanns24,
  title = {Predicting correlations in superradiant emission from a cascaded quantum system},
  author = {Tebbenjohanns, Felix and Mink, Christopher D. and Bach, Constanze and Rauschenbeutel, Arno and Fleischhauer, Michael},
  journal = {Phys. Rev. A},
  volume = {110},
  issue = {4},
  pages = {043713},
  numpages = {12},
  year = {2024},
  month = {Oct},
  publisher = {American Physical Society},
  doi = {10.1103/PhysRevA.110.043713},
  url = {https://link.aps.org/doi/10.1103/PhysRevA.110.043713}
}

@article{Anschuetz2023,
   title={Efficient classical algorithms for simulating symmetric quantum systems},
   volume={7},
   ISSN={2521-327X},
   url={http://dx.doi.org/10.22331/q-2023-11-28-1189},
   DOI={10.22331/q-2023-11-28-1189},
   journal={Quantum},
   publisher={Verein zur Forderung des Open Access Publizierens in den Quantenwissenschaften},
   author={Anschuetz, Eric R. and Bauer, Andreas and Kiani, Bobak T. and Lloyd, Seth},
   year={2023},
   month=nov, pages={1189} }

@misc{Goh2023,
      title={Lie-algebraic classical simulations for variational quantum computing}, 
      author={Matthew L. Goh and Martin Larocca and Lukasz Cincio and M. Cerezo and Fr\'ed\'eric Sauvage Sauvage},
      year={2023},
      eprint={2308.01432},
      archivePrefix={arXiv},
      primaryClass={quant-ph},
      url={https://arxiv.org/abs/2308.01432}, 
}

@article{Rand1988,
title = {On the identification of a {L}ie algebra given by its structure constants. {I}. {D}irect decompositions, levi decompositions, and nilradicals},
journal = {Linear Algebra and its Applications},
volume = {109},
pages = {197-246},
year = {1988},
issn = {0024-3795},
doi = {https://doi.org/10.1016/0024-3795(88)90210-8},
url = {https://www.sciencedirect.com/science/article/pii/0024379588902108},
author = {D. Rand and P. Winternitz and H. Zassenhaus}
}

@article{Lukin23,
  title = {Two-Emitter Multimode Cavity Quantum Electrodynamics in Thin-Film Silicon Carbide Photonics},
  author = {Lukin, Daniil M. and Guidry, Melissa A. and Yang, Joshua and Ghezellou, Misagh and Deb Mishra, Sattwik and Abe, Hiroshi and Ohshima, Takeshi and Ul-Hassan, Jawad and Vu\ifmmode \check{c}\else \v{c}\fi{}kovi\ifmmode \acute{c}\else \'{c}\fi{}, Jelena},
  journal = {Phys. Rev. X},
  volume = {13},
  issue = {1},
  pages = {011005},
  numpages = {17},
  year = {2023},
  month = {Jan},
  publisher = {American Physical Society},
  doi = {10.1103/PhysRevX.13.011005},
  url = {https://link.aps.org/doi/10.1103/PhysRevX.13.011005}
}

@article{Carmichael00,
title = {A quantum trajectory unraveling of the superradiance master equation},
journal = {Optics Communications},
volume = {179},
number = {1},
pages = {417-427},
year = {2000},
issn = {0030-4018},
doi = {https://doi.org/10.1016/S0030-4018(99)00694-X},
url = {https://www.sciencedirect.com/science/article/pii/S003040189900694X},
author = {H.J. Carmichael and Kisik Kim},
}

@article{Vetsch10,
	Author = {Vetsch, E. and Reitz, D. and Sagu\'e, G. and Schmidt, R. and Dawkins, S. T. and Rauschenbeutel, A.},
	Doi = {10.1103/PhysRevLett.104.203603},
	Issue = {20},
	Journal = {Phys. Rev. Lett.},
	Month = {May},
	Numpages = {4},
	Pages = {203603},
	Publisher = {American Physical Society},
	Title = {Optical Interface Created by Laser-Cooled Atoms Trapped in the Evanescent Field Surrounding an Optical Nanofiber},
	Url = {https://link.aps.org/doi/10.1103/PhysRevLett.104.203603},
	Volume = {104},
	Year = {2010}}

@article{Goban12,
	Author = {Goban, A. and Choi, K. S. and Alton, D. J. and Ding, D. and Lacro\^ute, C. and Pototschnig, M. and Thiele, T. and Stern, N. P. and Kimble, H. J.},
	Doi = {10.1103/PhysRevLett.109.033603},
	Issue = {3},
	Journal = {Phys. Rev. Lett.},
	Month = {Jul},
	Numpages = {5},
	Pages = {033603},
	Publisher = {American Physical Society},
	Title = {Demonstration of a State-Insensitive, Compensated Nanofiber Trap},
	Url = {https://link.aps.org/doi/10.1103/PhysRevLett.109.033603},
	Volume = {109},
	Year = {2012}}

@article{Gouraud15,
	Author = {Gouraud, B. and Maxein, D. and Nicolas, A. and Morin, O. and Laurat, J.},
	Doi = {10.1103/PhysRevLett.114.180503},
	Issue = {18},
	Journal = {Phys. Rev. Lett.},
	Month = {May},
	Numpages = {5},
	Pages = {180503},
	Publisher = {American Physical Society},
	Title = {Demonstration of a Memory for Tightly Guided Light in an Optical Nanofiber},
	Url = {https://link.aps.org/doi/10.1103/PhysRevLett.114.180503},
	Volume = {114},
	Year = {2015}}

@article{Solano17,
	Author = {Solano, P. and Barberis-Blostein, P. and Fatemi, F. K. and Orozco, L. A. and Rolston, S. L.},
	Da = {2017/11/30},
	Doi = {10.1038/s41467-017-01994-3},
	Id = {Solano2017},
	Isbn = {2041-1723},
	Journal = {Nat. Commun.},
	Number = {1},
	Pages = {1857},
	Title = {Super-radiance reveals infinite-range dipole interactions through a nanofiber},
	Volume = {8},
	Year = {2017}}

@article{Lodahl15,
  title = {Interfacing single photons and single quantum dots with photonic nanostructures},
  author = {Lodahl, Peter and Mahmoodian, Sahand and Stobbe, S\o{}ren},
  journal = {Rev. Mod. Phys.},
  volume = {87},
  issue = {2},
  pages = {347--400},
  numpages = {54},
  year = {2015},
  month = {May},
  publisher = {American Physical Society},
  doi = {10.1103/RevModPhys.87.347},
  url = {https://link.aps.org/doi/10.1103/RevModPhys.87.347}
}

@article{
Tiranov23,
author = {Alexey Tiranov  and Vasiliki Angelopoulou  and Cornelis Jacobus van Diepen  and Björn Schrinski  and Oliver August Dall’Alba Sandberg  and Ying Wang  and Leonardo Midolo  and Sven Scholz  and Andreas Dirk Wieck  and Arne Ludwig  and Anders S\o{}ndberg S\o{}rensen  and Peter Lodahl },
title = {Collective super- and subradiant dynamics between distant optical quantum emitters},
journal = {Science},
volume = {379},
number = {6630},
pages = {389-393},
year = {2023},
doi = {10.1126/science.ade9324},}

@article{Mirhosseini19,
	Author = {Mirhosseini, Mohammad and Kim, Eunjong and Zhang, Xueyue and Sipahigil, Alp and Dieterle, Paul B. and Keller, Andrew J. and Asenjo-Garcia, Ana and Chang, Darrick E. and Painter, Oskar},
	Da = {2019/05/01},
	Doi = {10.1038/s41586-019-1196-1},
	Id = {Mirhosseini2019},
	Isbn = {1476-4687},
	Journal = {Nature},
	Number = {7758},
	Pages = {692--697},
	Title = {Cavity quantum electrodynamics with atom-like mirrors},
	Ty = {JOUR},
	Url = {https://doi.org/10.1038/s41586-019-1196-1},
	Volume = {569},
	Year = {2019}}

@Article{Zanner22,
author={Zanner, Maximilian
and Orell, Tuure
and Schneider, Christian M. F.
and Albert, Romain
and Oleschko, Stefan
and Juan, Mathieu L.
and Silveri, Matti
and Kirchmair, Gerhard},
title={Coherent control of a multi-qubit dark state in waveguide quantum electrodynamics},
journal={Nature Physics},
year={2022},
month={May},
day={01},
volume={18},
number={5},
pages={538-543},
url={https://doi.org/10.1038/s41567-022-01527-w}
}

@misc{SI,
  note = {See Supplemental Information at [URL] for further details on the derivation of the ring resonator {G}reen's function, the identification of the canonical basis for exploiting symmetry via algebraic methods, the characterization of dark states with $kd=2\pi/3$, the extension to local dissipative processes, and a study of robustness, which includes Refs. \cite{Rand1988,Tai, AnaPRA, AnaPRX, Pineiro22}},
  presort = {mm},
}

@article{Schafer24,
  author={V. M. Sch{\"a}fer and Z. Niu and J. R. K. Cline and D. J. Young and E. Y. Song and H. Ritsch and J. K. Thompson},
  title     = {Continuous recoil-driven lasing and cavity frequency pinning with laser-cooled atoms},
  journal   = {Nature Physics},
  volume    = {21},
  pages     = {902--908},
  year      = {2025},
  doi       = {10.1038/s41567-025-02854-4},
  url       = {https://www.nature.com/articles/s41567-025-02854-4}
}

@article{Pineiro22,
  title = {Emergent Dark States from Superradiant Dynamics in Multilevel Atoms in a Cavity},
  author = {Pi\~neiro Orioli, A. and Thompson, J. K. and Rey, A. M.},
  journal = {Phys. Rev. X},
  volume = {12},
  issue = {1},
  pages = {011054},
  numpages = {36},
  year = {2022},
  month = {Mar},
  publisher = {American Physical Society},
  doi = {10.1103/PhysRevX.12.011054},
  url = {https://link.aps.org/doi/10.1103/PhysRevX.12.011054}
}

@article{Shankar22,
  title = {Simulating Dynamical Phases of Chiral $p+ip$ Superconductors with a Trapped ion Magnet},
  author = {Shankar, Athreya and Yuzbashyan, Emil A. and Gurarie, Victor and Zoller, Peter and Bollinger, John J. and Rey, Ana Maria},
  journal = {PRX Quantum},
  volume = {3},
  issue = {4},
  pages = {040324},
  numpages = {23},
  year = {2022},
  month = {Nov},
  publisher = {American Physical Society},
  doi = {10.1103/PRXQuantum.3.040324},
  url = {https://link.aps.org/doi/10.1103/PRXQuantum.3.040324}
}

@article{Eberly71,
  title = {Superradiance},
  author = {Rehler, Nicholas E. and Eberly, Joseph H.},
  journal = {Phys. Rev. A},
  volume = {3},
  issue = {5},
  pages = {1735--1751},
  numpages = {0},
  year = {1971},
  month = {May},
  publisher = {American Physical Society},
  doi = {10.1103/PhysRevA.3.1735},
  url = {https://link.aps.org/doi/10.1103/PhysRevA.3.1735}
}

@book{GoodmanBook,
  title     = "Invariants of the Classical Groups",
  author    = "Goodman, Roe and Wallach, Nolan R.",
  year      = 1998,
  publisher = "Cambridge University Press",
  address   = ""
}

@article{Manzoni17,
	title={{Simulating quantum light propagation through atomic ensembles using matrix product states}},
	author={Marco T. Manzoni and Darrick E. Chang and James S. Douglas},
	journal={Nature Communications},
	volume = {1743},
        issue = {8},
	year={2017},
	publisher={Nature},
	doi={10.1038/s41467-017-01416-4},
	url={https://www.nature.com/articles/s41467-017-01416-4},
}

@misc{Malz2024,
      title={Efficient tensor network simulation of multi-emitter non-{M}arkovian systems}, 
      author={Irene Papaefstathiou and Daniel Malz and J. Ignacio Cirac and Mari Carmen Bañuls},
      year={2024},
      eprint={2407.10140},
      archivePrefix={arXiv},
      primaryClass={quant-ph},
      url={https://arxiv.org/abs/2407.10140}, 
}

@article{Verstraete04,
  title = {Matrix Product Density Operators: Simulation of Finite-Temperature and Dissipative Systems},
  author = {Verstraete, F. and Garc\'{\i}a-Ripoll, J. J. and Cirac, J. I.},
  journal = {Phys. Rev. Lett.},
  volume = {93},
  issue = {20},
  pages = {207204},
  numpages = {4},
  year = {2004},
  month = {Nov},
  publisher = {American Physical Society},
  doi = {10.1103/PhysRevLett.93.207204},
  url = {https://link.aps.org/doi/10.1103/PhysRevLett.93.207204}
}

@article{AnaPRX,
  title = {Exponential Improvement in Photon Storage Fidelities Using Subradiance and ``Selective Radiance'' in Atomic Arrays},
  author = {Asenjo-Garcia, A. and Moreno-Cardoner, M. and Albrecht, A. and Kimble, H. J. and Chang, D. E.},
  journal = {Phys. Rev. X},
  volume = {7},
  issue = {3},
  pages = {031024},
  numpages = {36},
  year = {2017},
  month = {Aug},
  publisher = {American Physical Society},
  doi = {10.1103/PhysRevX.7.031024},
  url = {https://link.aps.org/doi/10.1103/PhysRevX.7.031024}
}

@article{Plankensteiner22,
  doi = {10.22331/q-2022-01-04-617},
  url = {https://doi.org/10.22331/q-2022-01-04-617},
  title = {Quantum{C}umulants.jl: {A} {J}ulia framework for generalized mean-field equations in open quantum systems},
  author = {Plankensteiner, David and Hotter, Christoph and Ritsch, Helmut},
  journal = {{Quantum}},
  issn = {2521-327X},
  publisher = {{Verein zur F{\"{o}}rderung des Open Access Publizierens in den Quantenwissenschaften}},
  volume = {6},
  pages = {617},
  month = jan,
  year = {2022}
}

@misc{Suresh25,
      title={Collective emission and selective-radiance in atomic clouds and arrays coupled to a microring resonator}, 
      author={Deepak A. Suresh and Xinchao Zhou and Chen-Lung Hung and F. Robicheaux},
      year={2025},
      eprint={2503.21121},
      archivePrefix={arXiv},
      primaryClass={quant-ph},
      url={https://arxiv.org/abs/2503.21121}, 
}

@misc{Zhou25,
      title={Selective collective emission from a dense atomic ensemble coupled to a nanophotonic resonator}, 
      author={Xinchao Zhou and Deepak A. Suresh and F. Robicheaux and Chen-Lung Hung},
      year={2025},
      eprint={2503.05664},
      archivePrefix={arXiv},
      primaryClass={quant-ph},
      url={https://arxiv.org/abs/2503.05664}, 
}

@misc{Holzinger24,
      title={Symmetry based efficient simulation of dissipative quantum many-body dynamics in subwavelength quantum emitter arrays}, 
      author={Raphael Holzinger and Oriol Rubies-Bigorda and Susanne F. Yelin and Helmut Ritsch},
      year={2024},
      eprint={2409.02790},
      archivePrefix={arXiv},
      primaryClass={quant-ph},
      url={https://arxiv.org/abs/2409.02790}, 
}

@article{Waugh68,
  title = {Approach to High-Resolution nmr in Solids},
  author = {Waugh, J. S. and Huber, L. M. and Haeberlen, U.},
  journal = {Phys. Rev. Lett.},
  volume = {20},
  issue = {5},
  pages = {180--182},
  numpages = {0},
  year = {1968},
  month = {Jan},
  publisher = {American Physical Society},
  doi = {10.1103/PhysRevLett.20.180},
  url = {https://link.aps.org/doi/10.1103/PhysRevLett.20.180}
}

@article{Kraus08,
  title = {Preparation of entangled states by quantum {M}arkov processes},
  author = {Kraus, B. and B\"uchler, H. P. and Diehl, S. and Kantian, A. and Micheli, A. and Zoller, P.},
  journal = {Phys. Rev. A},
  volume = {78},
  issue = {4},
  pages = {042307},
  numpages = {9},
  year = {2008},
  month = {Oct},
  publisher = {American Physical Society},
  doi = {10.1103/PhysRevA.78.042307},
  url = {https://link.aps.org/doi/10.1103/PhysRevA.78.042307}
}

@article{Guardiola25,
  author  = {E. Guardiola-Navarrete and S. Cardenas-Lopez and A. Asenjo-Garcia},
  title   = {Phase-space methods for many-body quantum optics},
  journal = {Adv. At. Mol. Opt. Phys.},
  volume  = {74},
  pages   = {87--156},
  year    = {2025},
doi     = {10.1016/bs.aamop.2025.04.003},
  url     = {https://doi.org/10.1016/bs.aamop.2025.04.003}
}

@misc{Lukin25,
      title={Mesoscopic cavity quantum electrodynamics with phase-disordered emitters in a Kerr nonlinear resonator}, 
      author={Daniil M. Lukin and Bennet Windt and Miguel Bello and Dominic Catanzaro and Melissa A. Guidry and Eran Lustig and Souvik Biswas and Giovanni Scuri and Trung Kien Le and Joshua Yang and Arina A. Nikitina and Misagh Ghezellou and Hiroshi Abe and Takeshi Ohshima and Jawad Ul-Hassan and Jelena Vučković},
      year={2025},
      eprint={2504.09324},
      archivePrefix={arXiv},
      primaryClass={quant-ph},
      url={https://arxiv.org/abs/2504.09324}, 
}

\clearpage
\onecolumngrid
\setcounter{equation}{0}
\begin{center}
    {\large \bf{Supplemental Material for \\ 
    Exact many-body quantum dynamics in one-dimensional baths via collective spins
}}\\
    \vspace{0.2cm}
    {\small Joseph T. Lee$^{1}$, Silvia Cardenas-Lopez$^{1}$, Stuart J. Masson$^{2}$, Rahul Trivedi$^{3}$, Ana Asenjo-Garcia$^{1}$}\\
    \vspace{0.2cm}
    {\small 
    $^{1}$Department of Physics, Columbia University, New York, New York 10027, USA\\
    $^{2}$Department of Physics, University of South Florida, Tampa, Florida 33620, USA\\
    $^{3}$Max Planck Institute of Quantum Optics, Hans-Kopfermann-Str. 1, Garching 85748, Germany
    }
    \texttt{ana.asenjo@columbia.edu}
\end{center}

\newenvironment{secondabstract}{
  \small
  \normalfont
}{}

\begin{secondabstract} We perform the supplementary calculations in support of the main text. We first calculate the Green's function for the ring resonator. We comment on a procedure to unveil the algebraic structure of the jump operators and show how the algorithm in Ref.~\cite{Rand1988} can be used to calculate a canonical basis. Additionally, we determine some characteristics for the dark states in the case of $kd=2\pi/3$, as well as the suppression of the decay rate of Dicke states for specific conditions. Finally, we demonstrate how the superspins formalism can be adjusted to include collective processes that break total angular momentum conservation, and present numerical results illustrating the robustness of partial permutational symmetry under positional disorder of the qubits.
\end{secondabstract}

\section{1. Ring Resonator Green's Function}
We derive the Green's function for qubits coupled to a ring resonator. The resulting Green's function is also derived in Ref. \cite{Hummer2013}. 

We model the ring cavity as a 1D channel with periodic boundary conditions, forming a ring of circumference $L$. Positions along the ring are denoted by $z$. To compute the Green's function, we sum over all clockwise and counterclockwise propagation paths from a source point $z'$ to an observation point $z$. Light can travel from $z'$ to $z$ either directly, acquiring a phase factor  $e^{ik|z-z'|}$, or via the complementary path around the ring, with a phase $e^{ik(L - |z-z'|)}$. Each full round trip around the ring contributes an additional phase factor of $e^{ikL}$. The sum is therefore
\begin{equation}
    g(z,z') = B \sum_{n=0}^\infty (e^{ikL} )^n\biggr(e^{ik|z-z'|}+e^{ik(L - |z-z'|)} \biggr), 
\end{equation}
where $B$ is a constant to be determined. The scalar Green's function must satisfy the 1D Helmholtz equation,
\begin{equation}
\label{eqn:Helmholtz1D}
    \biggr(\frac{d^2}{dz^2} + k^2\biggr) g = -\delta(z-z').
\end{equation}
To determine $B$, we integrate Eq.~\eqref{eqn:Helmholtz1D} over the interval $[z'+\epsilon,z'-\epsilon]$, and obtain \cite{Tai}
\begin{equation*}
    \lim_{\epsilon \rightarrow 0}\biggr(\biggr[ \frac{dg}{dz}\biggr]_{z'-\epsilon}^{z'+\epsilon} + k^2\int_{z'-\epsilon}^{z'+\epsilon} dz ~g\biggr)= -1.
\end{equation*}
Since $g(z,z')$ must be finite, the second integral vanishes in the limit $\epsilon \rightarrow 0$. This yields $B = \frac{i}{2k}$, resulting in the scalar Green's function
\begin{equation}
    g(z,z') =\frac{i}{2k} \sum_{n=0}^\infty (e^{ikL} )^n\biggr(e^{ik|z-z'|}+e^{ik(L - |z-z'|)} \biggr)=\frac{ic}{2\omega} \frac{1}{1 - e^{ikL}} \biggr( e^{ik|z-z'|} + e^{ikL} e^{-ik|z-z'|} \biggr),
\end{equation}
where we have rewritten the prefactor as $\frac{i}{2k} = \frac{ic}{2\omega}$. We express the Green's function in the more familiar frequency representation by considering the case close to resonance, where the wavevector is $k = k_r -\Delta + i \delta k$, with $\delta k$ and $\Delta$ real, and $k_r L = 2\pi m$ with $m$ an integer. If the losses from a single round trip can be neglected, or if $\delta k L, \Delta L \ll 1$, then
\begin{equation}
    e^{ik|z-z'|} + e^{ikL} e^{-ik|z-z'|} \approx 2\cos(k_r[z-z']).
\end{equation}
We can show $\cos(k[z-z']) \approx \cos(k_r[z-z'])$ by applying the double angle formula and removing terms that are first order in $\delta k$ and $\Delta$. Plugging in $k = k_r +\Delta_k  + i\delta k$ for the remaining parts of the Green's function yields 
\begin{equation}
    g(z,z',\omega) \approx \frac{ic}{\omega} \frac{1}{1 - e^{+i\Delta_k L}e^{-\delta k L}} \cos(k_r[z-z']).
\end{equation}
Let $e^{-\delta k L} = \bar{r}^2$. Following the derivation from \cite{AnaPRA}, we observe that $1 - \bar{r}^2e^{+i\Delta_k L} \approx 1 - \bar{r}^2 - i\bar{r}^2 \Delta_k L.$ Therefore,
\begin{equation}
    g(z,z',\omega) \approx \frac{ic}{\omega} \frac{1}{1 - \bar{r}^2 - i\bar{r}^2 \Delta_k L} \cos(k_r[z-z']).
\end{equation}
We introduce the notation $(1-\bar{r}^2)c/L = \kappa_c/2$. Additionally, let $\Delta_\omega = \bar{r}^2c\Delta_k = \bar{r}^2c\cdot \text{Re}(k - k_r) = \bar{r}^2\text{Re}(\omega - \omega_r)$ (thus having $k = \omega / c$). Simplifying yields our final Green's function to be
\begin{equation}
   g(z,z',\omega)=\frac{-c^2}{\omega L}  \frac{1}{\Delta_\omega + i \kappa_c/2 }\cos(k_r[z-z']). 
\end{equation}
The Green's function can thus be expressed as
\begin{equation}
    G(z,z',\omega)=\frac{-Ac^2}{\omega L}  \frac{1 }{\Delta_\omega + i \kappa_c/2}\cos(k_r[z-z']),
\end{equation}
where $A$ is the effective mode area. This factor emerges due to the model going from a three-dimensional problem to a one-dimensional Green's function \cite{AnaPRA}. The coherent evolution that comes from the real part of the Green's function vanishes when $\Delta_\omega$ is small. The dissipative dynamics are governed by the couplings $\Gamma^{ij}$. This can be obtained by taking the imaginary part \cite{AnaPRX}
\begin{equation}
    \Gamma^{ij} = \frac{2\mu_0 \omega_0^2}{\hbar} \wp^* \cdot \text{Im } \pmb{G}(z_i, z_j, \omega_0) \cdot \wp=\ga \cos(k_r[z_j-z_j]),
\end{equation}

with $\ga$ being  the individual decay rate into the ring cavity. At particular $k_r$, this allows for partial permutational symmetry among qubits.

\section{2. Lie Algebra of the Jump Operators}
The dissipative part of the evolution [Eq.~(7) of the main text] can be expressed in terms of different collective jump operators. The procedure to uncover the algebraic structure described in the main text is simplified by working in the directional basis,
\begin{equation}\label{LROps}
    \hat{\mathcal{O}}_{L(R)} = \frac{1}{\sqrt{N}} \sum_{j=1}^N e^{+(-) ikdj}\hat{\sigma}_-^j,
\end{equation}
which corresponds to photon emission into the left- (right-) propagating mode, respectively.

To unveil the algebraic structure, we compute the operators generated by the jump operators and their adjoints, $\{ \hat{\mathcal{O}}_{L},\hat{\mathcal{O}}_{R},\hat{\mathcal{O}}^\dagger_{L},\hat{\mathcal{O}}^\dagger_{R}\}$. Commutators of the directional operators yield raising, lowering, and $\hat{\sigma}_z$-type operators. For example,
\begin{equation}
    [\hat{\mathcal{O}}_\alpha^\dagger,  \hat{\mathcal{O}}_\beta]= \frac{1}{N} \sum_{j=1}^N e^{2ikdj\phi_{\alpha\beta}}\hat{\sigma}_z^j.
\end{equation}
where $\phi_{\alpha\beta}=1(-1)$ for $\{\alpha\beta\}=\{R,L\} (\{L,R\})$ and $\phi_{\alpha\beta}=0$ if $\alpha=\beta$. The coefficients on the new $\hat{\sigma}_z$-type operator take the form of complex phases. Taking commutators with the jump operators and $\hat{\sigma}_z$-type operators yield more jump operators that bear similarity to the structure of ladder operators. In fact, further commutators among the elements of the algebra either reproduce existing operators or generate new operators of the form
\begin{align}
    \hat{\mathcal{O}}_{\pm t} &= \frac{1}{\sqrt{N}}\sum_{j=1}^N e^{\pm  ikdtj}\hat{\sigma}_-^j, \\
   \hat{\mathcal{O}}_{z,\pm s} &= \frac{1}{N}\sum_{j=1}^N e^{\pm  ikdsj}\hat{\sigma}_z^j,
\end{align}
where $t$ iterates over odd integers and $s$ iterates over even integers. It can then be seen that the commutators generate a finite number of terms if the powers $e^{\pm  ikdtj}$ and $e^{\pm  ikdsj}$ eventually repeat, which occurs at rational fraction spacings. Specifically, for spacings $kd = n\pi/p$ with $n, p \in \mathbb{Z}^+$, the resulting algebra can be generated by only $3p$ distinct operators, which we denote as $\mathcal A_{n,p}$.

\subsection{2.1 Finding the Canonical Basis for $kd=\pi/2$}
The algebra obtained through the procedure described above has the same dimension as that of the $p$ angular momentum operators used in the superspin basis. However, the resulting operators do not coincide with those of Eq.~(5) in the main text. The canonical basis of angular momentum operators can be recovered from $\mathcal A_{n,p}$ by following the algorithm presented in Ref.~\cite{Rand1988}. As an illustrative example, we consider the case $kd=\pi/2$, for which the algebra consists of six operators, $\{ \hat{\mathcal{O}}_R^\dagger, \hat{\mathcal{O}}_R, \hat{\mathcal{O}}_L^\dagger, \hat{\mathcal{O}}_L, \hat{\mathcal{O}}_{z,0}, \hat{\mathcal{O}}_{z,2}\}$. In brief, the procedure involves finding the centralizer of the adjoint representation, computing the minimal polynomial, and determining the diagonalization matrix $G$ which transforms each element $e_i\in \mathcal A_{n,p}$ into a canonical basis element via $e_i'=Ge_i$.

We first compute the adjoint representation of the elements of the algebra, defined as
\begin{equation*}
     \text{ad}(e_i)\cdot e_j=\left[e_i,e_j\right].
\end{equation*}
This representation can be rewritten as matrices. The entries are given by $i$ and $j$ in a given sorted order of the basis. For $\mathcal{A}_{1,2}$ we have
\begin{align*}
    \text{ad}(\hat{\mathcal{O}}_R^\dagger) = \begin{pmatrix}0 & 0 & 0 & 0 & 0 & 0 \\ 0 & 0 & 0 & 0 & 1 & 0 \\ 0 & 0 & 0 & 0 & 0 & 0 \\ 0 & 0 & 0 & 0 & 0 & 1 \\ -2/N & 0 & 0 & 0 & 0 & 0 \\ 0 & 0 & -2/N & 0 & 0 & 0\end{pmatrix},~~ \text{ad}(\hat{\mathcal{O}}_R) = \begin{pmatrix}0 & 0 & 0 & 0 & -1 & 0 \\ 0 & 0 & 0 & 0 & 0 & 0 \\ 0 & 0 & 0 & 0 & 0 & -1 \\ 0 & 0 & 0 & 0 & 0 & 0 \\ 0 & 2/N & 0 & 0 & 0 & 0 \\ 0 & 0 & 0 & 2/N & 0 & 0\end{pmatrix},\\
    \text{ad}(\hat{\mathcal{O}}_L^\dagger) = \begin{pmatrix}0 & 0 & 0 & 0 & 0 & 0 \\ 0 & 0 & 0 & 0 & 0 & 1 \\ 0 & 0 & 0 & 0 & 0 & 0 \\ 0 & 0 & 0 & 0 & 1 & 0 \\ 0 & 0 & -2/N & 0 & 0 & 0 \\ -2/N & 0 & 0 & 0 & 0 & 0\end{pmatrix},~~ \text{ad}(\hat{\mathcal{O}}_L) = \begin{pmatrix}0 & 0 & 0 & 0 & 0 & -1 \\ 0 & 0 & 0 & 0 & 0 & 0 \\ 0 & 0 & 0 & 0 & -1 & 0 \\ 0 & 0 & 0 & 0 & 0 & 0 \\ 0 & 0 & 0 & 2/N & 0 & 0 \\ 0 & 2/N & 0 & 0  & 0 & 0\end{pmatrix},\\
    \text{ad}(\hat{\mathcal{O}}_{z,0}) = \begin{pmatrix}2/N & 0 & 0 & 0 & 0 & 0 \\ 0 & -2/N & 0 & 0 & 0 & 0 \\ 0 & 0 & 2/N & 0 & 0 & 0 \\ 0 & 0 & 0 & -2/N & 0 & 0 \\ 0 & 0 & 0 & 0 & 0 & 0 \\ 0 & 0 & 0 & 0 & 0 & 0\end{pmatrix},~~ \text{ad}(\hat{\mathcal{O}}_{z,2}) = \begin{pmatrix}0 & 0 & 2/N & 0 & 0 & 0 \\ 0 & 0 & 0 & -2/N & 0 & 0 \\ 2/N & 0 & 0 & 0 & 0 & 0 \\ 0 & -2/N & 0 & 0 & 0 & 0 \\ 0 & 0 & 0 & 0 & 0 & 0 \\ 0 & 0 & 0 & 0  & 0 & 0\end{pmatrix}.
\end{align*}

The centralizer, $C_R(\text{ad} (\mathcal{A}_{1,2}))$, is the set of operators that commute with all the Jordan normal forms of each of the elements of $\mathcal{A}_{1,2}$. While finding $C_R(\text{ad} (\mathcal{A}_{1,2}))$ can be computationally expensive in general, in this case the six operators yield only two distinct Jordan normal forms. We find that the centralizer includes the identity and

\begin{equation}
    a_0 = \begin{pmatrix} 0 & 1 & 0 & 0 & 0 & 0 \\ 0 & 0 & 1 & 0 & 0 & 0 \\ 1 & 0 & 0 & 0 & 0 & 0 \\ 0 & 1 & 0  & 0 & 0 & 0 \\ 0 & 0 & 0 & 0 & 0 & 1\\ 0 & 0 & 0 & 0 & 1 &0\end{pmatrix}.
\end{equation}
Next, we compute the characteristic polynomial of $a_0$. Then, computing the minimal polynomial and following the prescription to find the idempotent yields 
\begin{equation}
    E_1 = \frac{1}{2}\begin{pmatrix}1 & 0 & -1 & 0 & 0 & 0 \\ 0 & 1 & 0 & -1 & 0 & 0 \\ -1 & 0 & 1 & 0 & 0 & 0\\ 0 & -1 & 0 & 1 & 0 & 0 \\ 0 & 0 & 0 & 0 & 1 & -1 \\ 0 & 0 & 0 & 0 & -1 & 1\end{pmatrix}.
\end{equation}
We then find the matrix $G$ that diagonalizes $E_1$. This matrix is
\begin{equation}
    G = \begin{pmatrix}1 & 0 & -1 & 0 & 0 & 0 \\ 0 & 1 & 0 & -1 & 0 & 0 \\ 0 & 0 & 0 & 0 & 1 & -1 \\ 1 & 0 & 1 & 0 & 0 & 0 \\ 0 & 1 & 0 & 1 & 0 & 0 \\ 0 & 0 & 0 & 0 & 1 & 1\end{pmatrix}.
\end{equation}
Taking $e_i'=Ge_i$, we obtain a new basis
\begin{subequations}
\begin{align}
    e_1'&=e_1 - e_3 = \hat{\mathcal{O}}^\dagger_R - \hat{\mathcal{O}}^\dagger_L,\\
    e_2'&=e_2 - e_4 = \hat{\mathcal{O}}_R - \hat{\mathcal{O}}_L, \\
    e_3'&=e_5 - e_6 = \hat{\mathcal{O}}_{z,0} - \hat{\mathcal{O}}_{z,2},\\
   e_4'&=e_1 + e_3 = \hat{\mathcal{O}}^\dagger_R +\hat{\mathcal{O}}^\dagger_L,\\
    e_5'&=e_2 + e_4 = \hat{\mathcal{O}}_R + \hat{\mathcal{O}}_L ,\\
    e_6'&=e_5 + e_6 = \hat{\mathcal{O}}_{z,0} + \hat{\mathcal{O}}_{z,2}.
\end{align}
\end{subequations}
Up to a factor, these six operators are equivalent to $\hat{J}_{2+}, \hat{J}_{2-}, \hat{J}_{2z}, \hat{J}_{1+}, \hat{J}_{1-}, \hat{J}_{1z}$, which is the canonical basis given by the collective spins.

\subsection{2.2 Lie Algebraic Approach to Spin Length}

The non-conservation of the total spin length $s$ [Fig.~2(b) of the main text] is a direct consequence of the system's underlying Lie subalgebra. For a superspin with $p\geq 2$, the algebra generated by the jump operators is not a single $\mathfrak{su}(2)$ but a direct sum of $p$ independent algebras, $\bigoplus_{a=1}^p \mathfrak{su}(2)_a$, with $3p$ total generators. The dynamics thus respects the decomposition of the Hilbert space $\mathcal{H}$ into $p$ independent subsystems, i.e.,
\begin{equation}\label{irreps}
    \mathcal{H} = \bigotimes_{a=1}^p \mathcal{H}_a =  \bigotimes_{a=1}^p\biggr( \bigoplus_{j_a = j_{a,\text{min}}}^{j_{a, \text{max}}} \mathcal{V}_{j_a} \otimes \mathcal{M}_{j_a} \biggr),
\end{equation}
where $\mathcal{V}_{j_a}$ is the spin-$j_a$ irrep of the $a$-th subsystem. If the initial state is fully inverted, the dynamics is constrained to the manifold with $j_a = n_a/2\; \forall \, a$.

This dynamical symmetry is incompatible with the conservation of the total collective spin $s$. A conserved $s$ would require the evolution to be confined to a single irreducible representation $\mathcal{V}_s$ within the collective spin decomposition
\begin{equation}
    \mathcal{H} = \bigoplus_{s=s_{\text{min}}}^{N/2} \mathcal{V}_{s} \otimes \mathcal{M}_{s},
\end{equation}
where $s = \{N/2, N/2-1,\dots\}$ labels the total spin irreps. However, the generators of the individual $\mathfrak{su}(2)_a$ algebras do not commute with the total spin Casimir operator $S^2$. Consequently, even when initialized in the pure $s=N/2$ irrep (the fully inverted state), the evolution governed by Eq.~\eqref{irreps} necessarily generates coherences between different collective $\mathcal{V}_s$ sectors. The total spin $s$ is therefore not a fixed quantity.

\section{3. Dark States}
\subsection{3.1 Dicke States as an approximation to the dark states}

In this section we first show that single excitation Dicke states are always dark for $kd = n\pi/p$ and arbitrary integers $n$ and $p$. We then demonstrate that the decay rate of Dicke states are suppressed by a factor of $\sim 1/N^2$ if qubits are arranged in superspin configurations provided that $n N/p$ is an integer even number.

 The single excitation Dicke states can be written as
\begin{equation}
    \ket{D_1^N} = \frac{1}{\sqrt{N}} \sum_{l_1 \in L_1^N} \ket{l_1},
\end{equation}

where we denote the state with the $l$th atom excited and the rest in the ground state by $\ket l$, and $L_1^N = \{1,2,\dots, N\}$, is the set of all possible string that indicate which qubit is excited. For example, we can write the 1 excitation, 3 qubit Dicke state as
\begin{equation}
    \ket{D_1^3} = \frac{1}{\sqrt{3}} \biggr(\ket{1} + \ket{2} + \ket{3}\biggr).
\end{equation}

Applying either the left or right jump operator in Eq.~\eqref{LROps}
yields
\begin{equation}
    \hat{\mathcal{O}}_{L(R)}\frac{1}{\sqrt{N}} \sum_{l_1 \in L_1^N} \ket{l_1} = \frac{1}{\sqrt{N}}\sum_{l_1 \in L_1^N} e^{+(-) ikd(l_1)}\ket{g}^{\otimes N},
\end{equation}

where $\ket{g}$ denotes the ground state. If $n N/p$ is an integer even number

\begin{equation}\label{sumPhases}
    \sum_{n=1}^Ne^{\pm ikdn}=0,
\end{equation}

so, unless $kd = 0 $ (which reduces to the Dicke limit), the single-excitation Dicke state is always dark.

We now consider the general case of $m$ excitations. We represent the $m$-excitation Dicke state of $N$ qubits as
\begin{equation}
    \ket{D_m^N} = \frac{1}{\sqrt{N \choose m}} \sum_{l_m^N \in L_m^N} \ket{l_m},
\end{equation}
where we sum over $L_m^N$, the set of all possible strings with $m$ unique numbers selected from 1 to $N$, and ${N \choose m}$ is the binomial coefficient that computes the number of combinations. Similarly to the single-excitation case, we denote by $\ket {l_m}$ the state with $m$ excited qubits, where the $\ell$th qubit is excited if $\ell\in l_m$, and it is in the ground state otherwise. For example, $L_2^3$ contains three strings: $\{1, 2\}, \{1, 3\},$ and $\{2, 3\}$, corresponding to $\ket{eeg}, \ket{ege},$ and $\ket{gee}$ respectively.

Applying the left and right jump operators, we have 
\begin{equation}
    \hat{{\mathcal{O}}}_{L(R)} \ket{D_m^N} =\frac{1}{\sqrt{N\cdot{N \choose m}}} \sum_{j=1}^N\sum_{l_m^N \in L_m^N}e^{+(-) ikdj}\hat{\sigma}_-^j \ket{l_m} \equiv\sum_{l_{m-1}^N \in L_{m-1}^N}\frac{\alpha(l_{m-1}^N)}{\sqrt{ N \cdot {N\choose m}}}\ket{l_{m-1}^N},
\end{equation}
where $\alpha(l_{m-1}^N)$ are coefficients that account for all the different decay processes that begin from a given $m$-excitation state and end in $\ket{l_{m-1}^N}$. Observe that 

$$e^{+(-) ikdj}\hat{\sigma}_-^j \ket{l_m}=\begin{cases}
    e^{+(-) ikdj}\ket{l_m/\{j\}}\: \:\:\:\text{  if } j\in l_m,\\
    0 \: \:\:\:\text{ otherwise }.
\end{cases}$$

This means that to compute the total phase contribution from all processes leading to a given state $\ket{l_{m-1}^N}$, we need only consider the strings obtained by adding one qubit to $l_{m-1}^N$. Summing over all such qubits not already in $l_{m-1}^N$, we find

\begin{equation}\label{alpha}
    \alpha(l_{m-1}^N) = \sum_{q \notin l_{m-1}^N} e^{+ikdq} = - \sum_{q \in l_{m-1}^N} e^{+ikdq},
\end{equation}

where the second equality follows from Eq.~\eqref{sumPhases}. The emission rate to the left is therefore given by
\begin{equation}
    R_L = \braket{D_m^N | \hat{{\mathcal{O}}}_L^\dagger\hat{{\mathcal{O}}}_L | D_m^N} = \sum_{l_{m-1}^N \in L_{m-1}^N}\frac{|\alpha(l_{m-1}^N)|^2 }{N \cdot {N\choose m}},
\end{equation}

or, plugging the expression for $\alpha(l_{m-1}^N)$ from Eq.~\eqref{alpha},

\begin{equation}
    R_L = \frac{1}{N \cdot {N\choose m}} \sum_{l_{m-1}^N \in L_{m-1}^N}^N \biggr| e^{+ikd l_{m-1}^N[1]} + e^{+ikd l_{m-1}^N[2]}+ \dots + e^{+ikd l_{m-1}^N[m-1]}\biggr|^2,
\end{equation}
where $l_{m-1}^N[i]$ denotes the $i$th entry in the string. The sum of phases inside the absolute value is bounded by $m-1$ (corresponding to perfect constructive interference). Since there are a total of  $N\choose m-1$ possible strings with $m-1$ elements, we have that the emission rate to the left is bounded by
\begin{equation}
    R_L \leq\frac{(m-1)^2}{N \cdot  {N\choose m}}  {N\choose m-1} = \frac{(m-1)^2 m}{N(N-m+1)}.
\end{equation}
We can find that the rate of emission to the right is bounded by the same quantity using identical arguments. Therefore, for low excitation number, the decay rate decreases as $\sim 1/N^2$ with qubit number.

\subsection{3.2 Dark states with $kd=2\pi/3$}
Here, we leverage symmetries to study the properties of the dark states accessible via collective decay for $kd=2\pi/3$ and $N$ divisible by 3, starting from the fully inverted state, $\ket{e}^{\otimes N}$. In this configuration, the system decomposes into three superspins, each a collective angular momentum with $j=\frac{1}{2}\frac{N}{3}$, capable of hosting up to $\frac{N}{3}$ excitations. As discussed in the main text, there exists an additional exchange symmetry among the three collective spins for this spacing. This symmetry becomes evident when the master equation is rewritten in terms of the collective spin operators,

\begin{align}
    \hat{\mathcal{L}}[\hat{\rho}] = \sum_{a,b=1}^p \frac{\tilde{\Gamma}^{ab}}{2} \biggr( 2\hat{J}_{b-} \hat{\rho} \hat{J}_{a+} - \{ \hat{J}_{a+}\hat{J}_{b-}, \hat{\rho}\}\biggr),
\end{align}
where $\Gamma^{ab} = \Gamma_{1D}$ for $a=b$ and $-\Gamma_{1D}/2$ for $a\neq b$. Note that exchanging collective spin indices leaves the master equation invariant.

Let $P( \ket{n_1, n_2, n_3})$ denote the exchange symmetric superposition of all states with $n_1, n_2, n_3$ excitations in the superspins 1,2 and 3, respectively. If the initial state is symmetric under superspin exchange -- for example, $\ket{\psi(0)}=\ket{e}^{\otimes N}$-- then the dynamics remains confined to this symmetric subspace.

In this section, we use the collective jump operators resulting from diagonalizing the $\tilde{\Gamma}$ matrix, rather than working in the directional basis in Eq.~\eqref{LROps} (although identical results would be obtained in the directional representation). We denote the new jump operators by $\{\hat{\mathcal{O}}_+,\hat{\mathcal{O}}_-\}$. For this particular case, they read
\begin{align}
    \hat{\mathcal{O}}_+ &= \sqrt{\frac{1}{6}} (-\hat{J}_{1-} + 2\hat{J}_{2-} - \hat{J}_{3-}),\\
    \hat{\mathcal{O}}_- &= \sqrt{\frac{1}{2}} (\hat{J}_{1-} - \hat{J}_{3-}).
\end{align}

To find the dark states with a fixed number of excitations that are accessible through the dynamics, we construct general symmetric states with that excitation number and impose that they are annihilated by $\hat{\mathcal{O}}_\pm$. For instance, in the two-excitation manifold we have

\begin{equation}
     \ket{2 \text{ ex.}} =A\cdot P(\ket{2, 0,0})+B\cdot P(\ket{1,1,0}),
\end{equation}
where $A,$ $B\in\mathbb C$ are two complex coefficients.  Imposing the condition $\hat{\mathcal{O}}_\pm \ket{2 \text{ ex.}} = 0$, we find a unique two-excitation dark state accessible from $\ket{e}^{\otimes N}$ under collective decay. Its explicit form is

\begin{equation}
     \ket{2 \text{ ex.}} =A\cdot P(\ket{2, 0,0})+\sqrt{2 - \frac{1}{j}}A\cdot P(\ket{1,1,0}),
\end{equation}
with $A = \sqrt{\frac{1}{3(3-1/j)}}$, fixed by the normalization condition. The procedure for finding dark states in other excitation manifolds is identical: imposing $\hat{\mathcal{O}}_\pm \ket{\psi} = 0$ leads to an algebraic system with an unique solution in each case. Thus, for every excitation, there is at most one dark state that can be populated by the dynamics starting from the fully inverted state.

Interestingly, no such dark state exists when the number of excitations exceeds $\frac{N}{3}$. We now prove this excitation limit. Since $P(\ket{n_1, n_2, n_3})$ is exchange symmetric, we can freely reorder the occupation numbers. We aim to show that if $\sum_a n_a = N/3+b$, with $b$ an integer in the range $[1,\frac{2 N}{3}]$, then no state that respects exchange symmetry among the superspins can be dark. To simplify the analysis, it is sufficient to consider the action of $\hat{\mathcal{O}}_{-}$, which acts only on the first and third superspin. We will show that it is impossible to construct a state with $N/3+b$ excitations that is dark with respect to this jump operator. We demonstrate this explicitly for the case $b=1$, noting that the argument extends identically to excitation manifolds with $b>1$.

The proof proceeds by contradiction. Suppose $\ket{\psi}$ is a dark state -- implying in particular that, $\hat{\mathcal{O}}_{-}\ket{\psi}=0$. We will show that in this case, no term of the form $\ket{n_1, n_2, n_3}$ can appear in its expansion, for any values of $n_a$. Since all basis elements are of this form, it follows that no such dark state exists. 

We prove that no $\ket{n_1, n_2, n_3}$ appears in $\ket{\psi}$ by induction. Let $n_1 = N/3 - q$. We begin with the base case, $q=0$. The symmetrized state $P(\ket{N/3, n_2, n_3})$ with $n_2+n_3=1$ cannot appear in the expansion of $\ket \psi$, as it would contribute with a term $\ket{N/3, n_2, n_3-1}$ to $\hat{\mathcal{O}}_{-}\ket \psi$. Crucially, the only other term that could yield $\ket{N/3, n_2, n_3-1}$ under the action of $\hat{\mathcal{O}}_{-}$ -- and thus potentially interfere destructively -- is $\ket{N/3+1, n_2, n_3}$, which is forbidden because each superspin can have at most $\frac{N}{3}$ excitations. Therefore, 
if $\ket{\psi}$ is dark, it cannot contain $P(\ket{N/3, 0, 1})$.
 
For the inductive step, assume that no term of the form $P(\ket{n_1, n_2, n_3})$ with $n_1 = N/3 - q$ appears in the expansion of $\ket{\psi}$. We aim to show that this implies the absence of the term $P(\ket{n_1= N/3 - (q+1), n_2, n_3})$ as well. 

Under the action of $\hat{\mathcal{O}}_{-}\ket \psi$, $P(\ket{n_1= N/3 - (q+1), n_2, n_3})$ contributes with the term $P( \ket{N/3 - (q+1), n_2, n_3-1})$. The only other term that could produce this same contribution is $P( \ket{N/3 - q, n_2, n_3-1})$, which by the induction hypothesis does not appear in $\ket \psi$. Since there is no other source that could interfere destructively with the contribution $P( \ket{N/3 - (q+1), n_2, n_3-1})$, we conclude that  $P(\ket{n_1= N/3 - (q+1), n_2, n_3})$ cannot appear in the expansion of $\ket{\psi}$.

Hence, no term of the form $P(\ket{n_1= N/3 - q, n_2, n_3})$ appears in the expansion of $\ket{\psi}$ for any $n_1\leq N/3$ in the manifold of $\frac{N}{3}+1$ excitations. Therefore, we conclude that no dark state exists in this manifold that also respects the additional exchange symmetry among superspins.

\subsection{3.3 Dark state calculation for $N=6$}
We explicitly calculate the two dark states for the case of $N=6$. We will use the basis $\ket{n_1,n_2,n_3}$ where for $N=6$ we have that $n_i = \{0, 1, 2\}$, corresponding to 0, 1, or 2 excitations in the ladder.

For the one excitation sector, we must build our states using $\ket{1, 0, 0}, \ket{0,1,0},$ and $\ket{0,0,1}$. While we can explictly calculate the states such that $\hat{\mathcal{O}}_\pm\ket{\psi}=0$, since we must have an additional exchange symmetry among the superspins, the only candidate for a dark state is
\begin{equation}
    \ket{\text{1 ex.}} = \frac{1}{\sqrt{3}}\biggr(\ket{1,0,0} + \ket{0,1,0}+\ket{0,0,1}\biggr).
\end{equation}
For the two excitation dark state, our basis elements will be
\begin{align}
    \ket{2, 0, 0} + \ket{0,2,0} + \ket{0,0,2},\\
    \ket{1, 1, 0} + \ket{1, 0, 1} + \ket{0, 1,1}.
\end{align}
We therefore need to find a dark state in the form 
\begin{equation}
    \ket{\text{2 ex.}} = A\biggr(\ket{2, 0, 0} + \ket{0,2,0} + \ket{0,0,2}\biggr) +B\biggr(\ket{1, 1, 0} + \ket{1, 0, 1} + \ket{0, 0,1}\biggr).
\end{equation}
The superspin operators acting on our basis states yield
\begin{align}
    \hat{J}_{1-}\ket{2,0,0} = \sqrt{2(2j-1)}\ket{1,0,0} = \sqrt{2}\ket{1,0,0}, \\
    \hat{J}_{1-}\ket{1,1,0}= \sqrt{2j}\ket{0,1,0} = \sqrt{2}\ket{0,1,0},
\end{align}
where we have used the angular momentum algebra and $j = N/(3\cdot 2) = 1$ to simplify. Acting with $\hat{\mathcal{O}}_+$ on the state, we obtain
\begin{align}
    \hat{\mathcal{O}}_+A(\ket{2, 0, 0} + \ket{0,2,0} + \ket{0,0,2}) &= A\frac{1}{\sqrt{3}}(-\ket{1, 0, 0} + 2 \ket{0,1,0} - \ket{0,0,1}),\\
    \hat{\mathcal{O}}_+B(\ket{1, 1, 0} + \ket{1, 0, 1} + \ket{0, 1,1}) &= B \frac{1}{\sqrt{3}}(-2\ket{0, 1, 0} + \ket{1, 0, 0}  +  \ket{0,0,1}).
\end{align}
Therefore, we find that $A = B.$ The general condition would yield $B = \sqrt{2 - 1/j}A$. Normalizing the dark state forces $|A|^2 = 6$, or
\begin{equation}
    \ket{\text{2 ex.}} = \frac{1}{\sqrt{6}}\biggr(\ket{2, 0, 0} + \ket{0,2,0} + \ket{0,0,2} +\ket{1, 1, 0} + \ket{1, 0, 1} + \ket{0, 0,1}\biggr).
\end{equation}

\subsection{3.4 Dark state populations }
\begin{figure}
\begin{center}
\includegraphics[width=9cm]{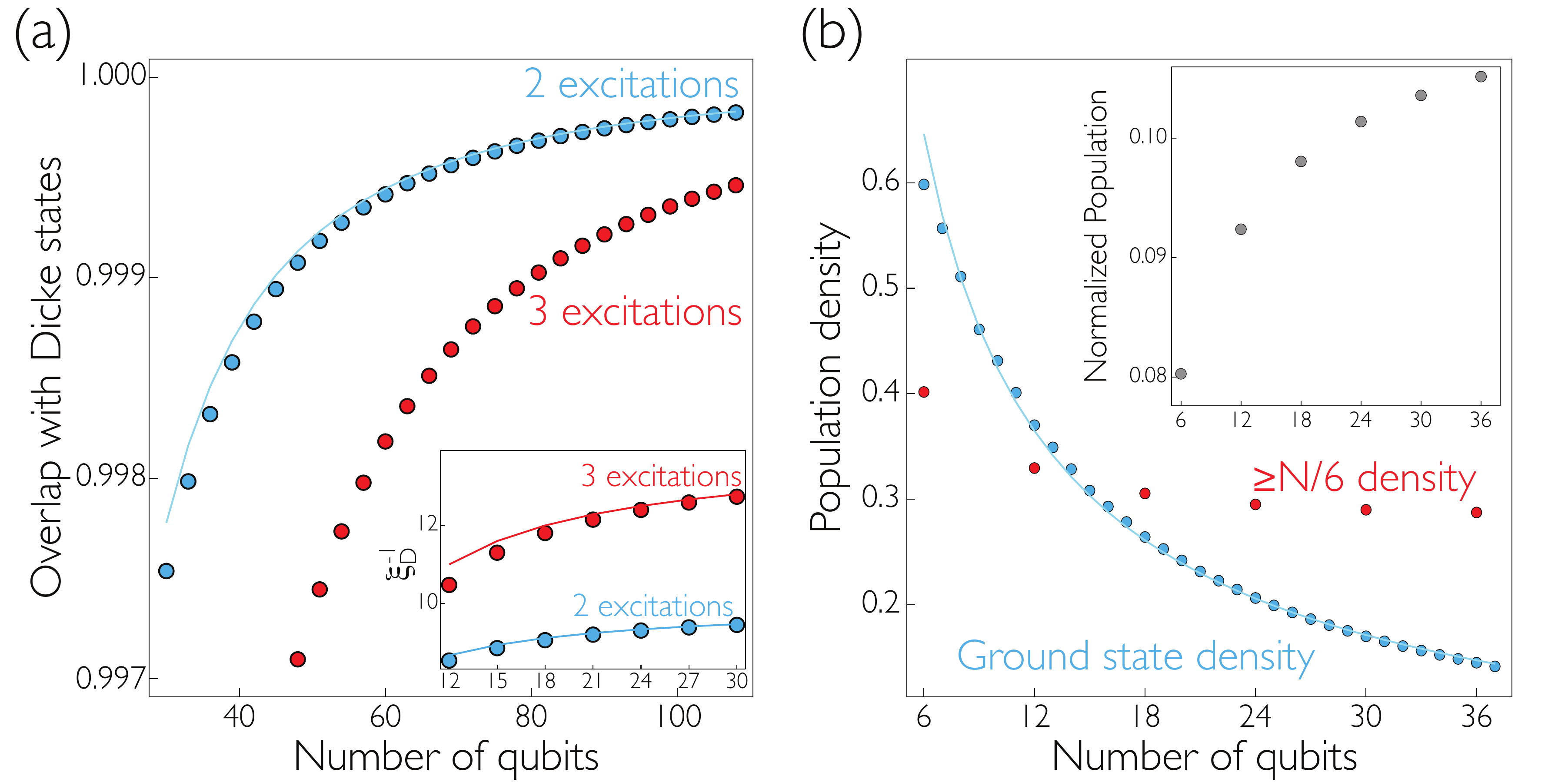}
\caption{\textbf{Entangled dark states for $kd=2\pi/3$ are unique and exhibit high overlap with fully-symmetric Dicke states.} (a)~Overlap between dark states and corresponding symmetric Dicke states for the two-excitation (blue) and three-excitation (red) manifolds. The solid line shows the analytical prediction for 2 excitations in the large $N$-limit, $F \approx 1 - 2/N^2$. Inset:~Inverse of the Dicke squeezing parameter $\xi_D^{-1}$ for the 2- and 3-excitation dark states, compared with symmetric Dicke states (solid lines). (b)~Population density of the ground state (blue), and of all states with at least $N/6$ excitations (red) as a function of qubit number $N$. Solid lines are power-law fits, scaling approximately as $\sim N^{-0.82}$ for the ground state. Inset: Normalized population for a given qubit number. Plots were obtained by numerically evolving to late times ($\Gamma_\text{1D}t=40$). The resulting states were numerically verified to be dark by computing the emission rates.}
\label{fig:2pi3-endmatter}
\end{center}
\end{figure}

For $kd=2\pi/3$ and $N$ divisible by 3, the dark states for each exitation manifold approaches the corresponding Dicke state. For example, the overlap (fidelity) between the two-excitation dark state $\ket{\psi}$ and a Dicke state is
\begin{equation}
    F=|\braket{\psi | N/2, -N/2 +2}|^2 = \frac{1}{1 + \frac{1}{9j(2j-1)}},
\end{equation}
where $j=\frac{1}{2}\frac{N}{3}$ denotes the total angular momentum of each superspin. In the $N\gg 1$ limit, the fidelity reduces to $F \approx 1 -2/N^2$, which agrees with numerical calculations, as shown in Fig.~\ref{fig:2pi3-endmatter}(a). Fully permutationally symmetric dark states are known to appear in multilevel atoms in a cavity~\cite{Pineiro22}. In our case, however, full permutational symmetry is not imposed by the model but emerges dynamically.

Additionally, decay dynamics from the fully inverted state can be used to generate metrologically useful dark states. As shown in Fig.~\ref{fig:2pi3-endmatter}(b), the ground state population decreases with increasing $N$, while states beyond the low excitation limit (e.g. with excitation number greater than $N/6$) retain non-negligible population.

\section{4. SU(4) solution for added local decay}
Realistic physical systems will have mechanisms that break the partial permutational symmetry. However, some of these mechanisms still allow for an efficient representation. For example, parasitic decay to modes other than the 1D environment is described by 
\begin{equation} 
     \hat{\mathcal{L}}_{\text{loc}}[\hat{\rho}] = \sum_{n=1}^N \frac{\Gamma'}{2} \biggr( 2\hat{\sigma}_-^n \hat{\rho} \hat{\sigma}_+^n - \{ \hat{\sigma}_+^n\hat{\sigma}_-^n, \hat{\rho}\}\biggr).
\end{equation} 
This decay breaks the symmetry term per term, so, for instance, quantum trajectories will populate states that no longer exhibit partial permutational symmetry. However, since the sum over all the local decay channels maintains the partial permutational symmetry, the symmetry is preserved at the density matrix level. Hence, evolution can still be captured in an efficient manner by using either Dicke states of different total angular momentum or SU(4) operators \cite{Molmer18, Holland13}. 

Explicitly, following the prescription in \cite{Holland13}, we use the same SU(4) superoperators. We assign one set of SU(4) superoperators for each superspin. In particular, we can show that for superspin $a$, 
\begin{subequations}
\begin{align}
    (U_-^a + V_-^a)\hat{\rho} &= \hat{\rho} \hat{J}_{a+},\\
    (U_+^a + V_+^a)\hat{\rho} &= \hat{\rho} \hat{J}_{a-},\\
    (M_-^a + N_-^a)\hat{\rho} &= \hat{J}_{a-} \hat{\rho}, \\
    (M_+^a + N_+^a)\hat{\rho} &= \hat{J}_{a+} \hat{\rho}.
\end{align}
\end{subequations}
We can write the dissipative evolution as
\begin{equation}
    \mathcal{L}{[\hat{\rho}]}= \sum_{ab}^p\Gamma^{ab} \biggr( (M_-^b + N_-^b) (U_-^a + V_-^a) - \frac{1}{2}(M_+^a + N_+^a)(M_-^b + N_-^b) - \frac{1}{2}(U_+^b + V_+^b)(U_-^a + V_-^a)\biggr) \hat{\rho}.
\end{equation}
Similarly, we can rewrite the local decay Lindbladian in terms of collective spin operators as
\begin{equation}
    \mathcal{L}_{loc}{[\hat{\rho}]}= \sum_{a=1}^p\Gamma_{loc} \biggr( Q_-^a - \frac{n_a}{2} - Q_3^a\biggr) \hat{\rho}.
\end{equation}
The resulting basis can be tracked with the quantum numbers $P(q^1, q_3^1, \sigma_3^1 ; q^2, q_3^2, \sigma_3^2; \dots ; q^p, q_3^p, \sigma_3^p  )$, where $q^a, q_3^a, \sigma_3^a$ are the quantum numbers for collective spin $a$. The new basis then scales as $\mathcal{O}((N/p)^{3p})$. A similar prescription can be used for other symmetry preserving mechanism such as certain pumping and dephasing mechanisms.

\section{5. Robustness}
Aspects that do not respect the partial permutational symmetry -- even at the level of the density matrix -- can cause the population to leak out of the subspace of states with efficient representations. Examples include frequency or positional disorder among qubits, or unequal coupling to the 1D environment, among others. A key question is how robust the superspin representation remains in the presence of such imperfections in the dynamics.

\begin{figure*}
\begin{center}
\includegraphics[width=17cm]{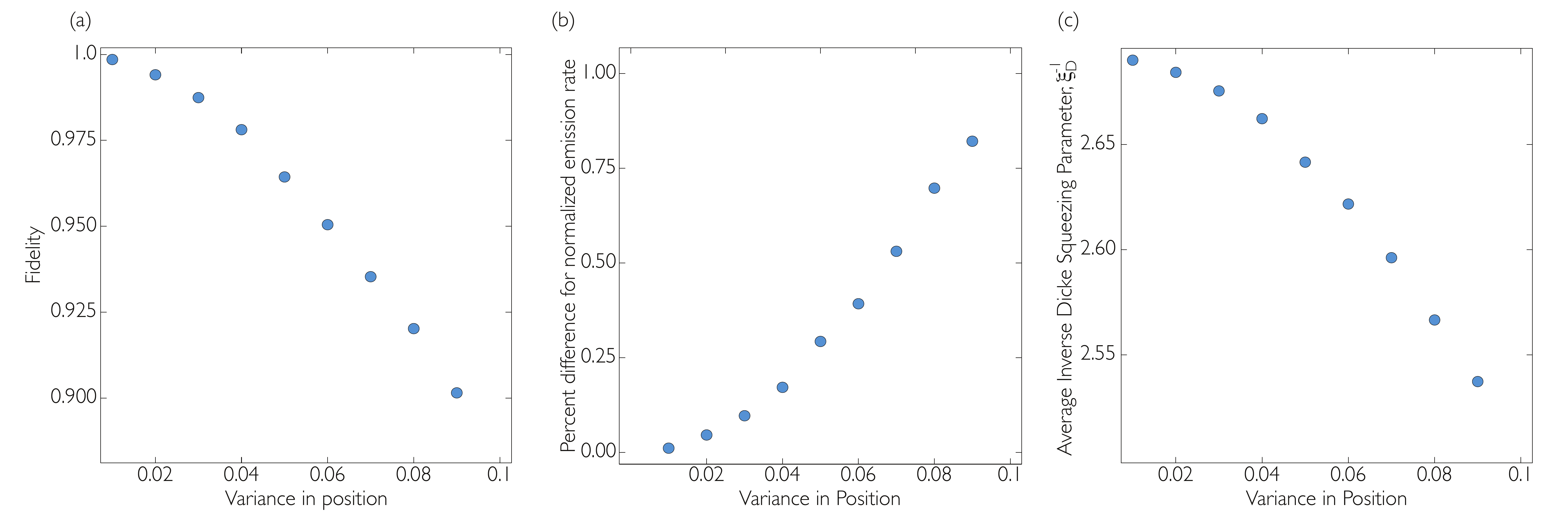}
\caption{Various results for 6 atoms at $kd=2\pi/3$ with variance in positions. The errors were drawn from a normal distribution with standard deviation $\sigma=\{0.01, 0.02, \dots, 0.09\}$. The new positions were changed from $kd_n = kdn$ to $kd_n'=kd(n + \epsilon_n)$. For all plots, the density matrix simulation was run 1000 times and averaged. (a)~Minimum fidelity of the perturbed spacing density matrices with the unperturbed states for $N=6$ qubits as a function of standard deviation. (b)~Percent difference from the non-disordered maximum emission rate, $R_{\text{max}, \sigma} / R_{\text{max}, \sigma = 0.0}$, plotted against the standard deviation. (c)~Average Inverse Dicke squeezing parameter $\xi_D^{-1}$ as a function of standard deviation.}
\label{fig:Robustness}
\end{center}
\end{figure*}

As an  example, we consider $6$ qubits decaying from the fully inverted state, and consider variances in the positions away from perfect rational spacings. Originally, the spacings are taken to be $kdn$ for the $n$th qubit. The perturbed spacings are given by $kd_n'=kd(n + \epsilon_n)$, where the errors $\epsilon_n$ are drawn from a normal distribution with variance $\sigma \in \{ 0.01, 0.02, \dots, 0.09\}$. 

We use the fidelity $\mathcal{F}(\hat{\rho}_1, \hat{\rho}_2) = ( tr\sqrt{\sqrt{\hat{\rho}_1}\hat{\rho}_2 \sqrt{\hat{\rho}_1}})^2$ averaged over several disorder realizations to study the effects of small perturbations to position. The minimum fidelity achieved through the decay decreases as expected, but for small variances, we maintain close to 90\% fidelity [Fig.~\ref{fig:Robustness}(a)]. 

We also examine the percent difference of the maximum emission normalized by the maximum emission rate of the non-perturbed configuration [Fig.~\ref{fig:Robustness}(b)]. As the variance increases, the maximum emission rate increases, but is still under 1\% greater than the non-disordered rate.

The average inverse Dicke Squeezing parameter decreases with the disorder~[Fig.~\ref{fig:Robustness}(c)]. We expect that small denominator spacings have maximum entanglement. Thus, increasing the disorder changes the configuration from perfectly ordered to random, leading to a drop in metrological advantage. 

\end{document}